\documentclass[useAMS,usenatbib,usegraphicx]{mn2e}
\usepackage{amssymb}
\usepackage{url}
\usepackage{mathptmx}

\include{defn}

\begin{document}

\newcommand{\FIXME}[1]{\textbf{FIXME: #1}}

\title{Probing the neutrino mass hierarchy with CMB weak lensing}
\author[Alex C. Hall, Anthony Challinor]
{Alex C. Hall,$^{1,2}$\thanks{ach74@ast.cam.ac.uk}Anthony Challinor$^{1,2,3}$\\
$^1$Institute of Astronomy, Madingley Road, Cambridge, CB3 0HA, U.K.\\
$^2$Kavli Institute for Cosmology Cambridge, Madingley Road, Cambridge, CB3 0HA, U.K.\\
$^3$DAMTP, Centre for Mathematical Sciences, Wilberforce Road, Cambridge, CB3 0WA, U.K.}
\maketitle

\begin{abstract}
We forecast constraints on cosmological parameters with primary CMB anisotropy information and weak lensing reconstruction with a future post-$Planck$ CMB experiment, the \emph{Cosmic Origins Explorer} (\textit{COrE}), using oscillation data on the neutrino mass splittings as prior information. Our MCMC simulations in flat models with a non-evolving equation-of-state of dark energy $w$ give typical 68\% upper bounds on the total neutrino mass of $0.136\,\mathrm{eV}$ and $0.098\,\mathrm{eV}$ for the inverted and normal hierarchies respectively, assuming the total summed mass is close to the minimum allowed by the oscillation data for the respective hierarchies ($0.10\,\mathrm{eV}$ and $0.06\,\mathrm{eV}$). Including geometric information from future baryon acoustic oscillation measurements with the complete BOSS, Type 1a supernovae distance moduli from \textit{WFIRST}, and a realistic prior on the Hubble constant, these upper limits shrink to $0.118\,\mathrm{eV}$ and $0.080\,\mathrm{eV}$ for the inverted and normal hierarchies, respectively. Addition of these distance priors also yields percent-level constraints on $w$. We find tension between our MCMC results and the results of a Fisher matrix analysis, most likely due to a strong geometric degeneracy between the total neutrino mass, the Hubble constant, and $w$ in the unlensed CMB power spectra. If the minimal-mass, normal hierarchy were realised in nature, the inverted hierarchy should be disfavoured by the full data combination at typically greater than the $2\sigma$ level. For the minimal-mass inverted hierarchy, we compute the Bayes' factor between the two hierarchies for various combinations of our forecast datasets, and find that the future cosmological probes considered here should be able to provide `strong' evidence (odds ratio 12:1) for the inverted hierarchy. Finally, we consider potential biases of the other cosmological parameters from assuming the wrong hierarchy and find that all biases on the parameters are below their $1\sigma$ marginalised errors.
\end{abstract}

\section{Introduction}
\label{sec:intro}

In the most recent extension to the Standard Model of particle physics, it has been established that at least two of the three neutrino mass eigenstates possess non-zero mass. The SuperKamiokande experiment provided the first evidence for this with the detection of flavour oscillations in atmospheric neutrinos~\citep{super-k_98,super-k_01,super-k_04}, and the phenomenon has since been observed in solar neutrinos by the Sudbury Neutrino Observatory~\citep{sno_01,sno_02}, in reactor anti-neutrinos at KamLAND~\citep{kamland_03}, and in accelerator neutrinos by K2K and MINOS~\citep{k2k_03,minos_08}.

The oscillation experiments not only reveal that neutrinos have mass, but that the three mass eigenstates have \emph{different} masses. Oscillations are only sensitive to differences in the squared masses of the three eigenstates, and not to the overall mass scale. We use the central values from the global fits in Appendix~B of the arXiv update to~\cite{maltoni_review_04}:
\begin{equation}
\begin{array}{l l}
m_2^2 - m_1^2 = 7.9^{+1.0}_{-0.8} \times 10^{-5} \mathrm{eV}^{2} \\
\\
| m_3^2 - m_1^2 | = 2.2^{+1.1}_{-0.8} \times 10^{-3} \rmn{eV}^2,
\end{array}
\label{eq:masssq}
\end{equation}
with $3\sigma$ confidence levels indicated.
Since the sign of $m_3^2 - m_1^2$ is unconstrained, there are two logical possibilities depending on the choice of sign. The difference is negligible for large total masses (the `degenerate' scenario, where $m_1 \sim m_2 \sim m_3$). At lower total masses, we have a hierarchical situation, with two distinct hierarchies demarcated by the oscillation data: `normal' ($m_1 < m_2 \ll m_3$) and `inverted' ($m_3 \ll m_1 < m_2$). Measuring the absolute mass scale and determining the true hierarchy of neutrinos are key issues in neutrino physics. Note that equation~(\ref{eq:masssq}) implies a lower limit on the total mass for each hierarchy: $0.095\,\mathrm{eV}$ for inverted, and $0.056\,\mathrm{eV}$ for normal.

Cosmology provides a different perspective on this problem, essentially by
being sensitive to the gravitational effect of neutrinos on the matter and radiation fields across cosmic time. The primary CMB temperature (T) and polarisation (P) anisotropies can go some way to constraining absolute neutrino masses, but are fairly insensitive to realistic (sub-eV) mass scales compared to the late-time influence of neutrino masses on the clustering of matter~\citep{elgaroy_05, ichikawa_05}. Tighter limits may therefore be obtained by including large scale structure information, such as the shape of the matter power spectrum and cosmic shear~\citep{1475-7516-2010-12-027}, as well as distance measures such as baryon acoustic oscillations (BAO) and Type 1a supernovae (see~\citealt{abazajianreview} for a recent review). With the seven-year \textit{WMAP} data alone, $\sum m_\nu < 1.3\,\mathrm{eV}$ (95\% C.L.) assuming a flat universe and fixed dark energy
equation of state $w=-1$~\citep{wmap7}; this improves to $\sum m_\nu < 0.58\,\mathrm{eV}$ when including BAO from SDSS DR7~\citep{percival_10} and a $H_0$ prior of 4\% width from~\citet{riess_09}. For the same model, the most recent constraints from adding galaxy clustering to WMAP are $\sum m_\nu < 0.36\,\mathrm{eV}$~\citep{2012arXiv1201.1909D}, using a photometric sample of luminous galaxies from SDSS DR8, and $\sum m_\nu < 0.51\,\mathrm{eV}$~\citep{2012arXiv1203.6616S}, using a spectroscopic galaxy sample from SDSS DR9 (as part of the ongoing Baryon Oscillation Spectroscopic Survey; BOSS).

Small-scale CMB measurements open up the possibility of exploiting the effect of weak gravitational lensing (WL) of the CMB as well as the primary anisotropies
(see~\citealt{anthonyreview} for a review of CMB lensing). CMB photons are deflected by large scale structure along the line of sight as they propagate to us from last-scattering, and these deflections can be reconstructed from their non-Gaussian imprint in the CMB at small angular scales~\citep{1999PhRvD..59l3507Z,2001ApJ...557L..79H,2002ApJ...574..566H,okamotoandhu}. CMB lens reconstruction thus provides both geometric and late-time clustering information `for free' and so additional sensitivity to neutrino masses. Since the source plane is essentially fixed -- the last scattering surface -- this method is free of uncertainties over source redshifts, as well as having the advantage of probing structure formation without the problems of bias and redshift-space distortions. Lens reconstruction has recently been used to measure the power spectrum of the deflection field with temperature data from the Atacama Cosmology Telescope (ACT; \citealt{2011PhRvL.107b1301D}) and the South Pole telescope (SPT; \citealt{2012arXiv1202.0546V}).


Lens reconstruction from the CMB temperature suffers from statistical noise due to chance correlations in the unlensed CMB that mimic the effect of lensing. This is such that temperature reconstructions will never supply cosmic-variance-limited measurements of the deflection power spectrum for multipoles $l > 100$. Polarization measurements are very helpful here~\citep{2002ApJ...574..566H}, since they intrinsically have more small-scale power and the $B$-mode of polarization is not confused by primary anisotropies. In principle, polarization can provide cosmic-variance limited reconstructions to multipoles $l \approx 500$, i.e.\ on all scales where linear theory applies. For this reason, lens reconstruction from polarization has become an important part of the science case for successors to the \textit{Planck} satellite, such as the proposed European-led \textit{Cosmic Origins Explorer} (\textit{COrE}; \citealt{core}), and the US-led \textit{CMBPol} (e.g.~\citealt{2008arXiv0805.4207B,2009arXiv0906.1188B}) and polarization upgrades to ACT~\citep{2010SPIE.7741E..51N} and SPT~\citep{2009AIPC.1185..511M}.

Structure formation is mostly sensitive to the summed neutrino mass, but does have some weak sensitivity to individual masses. This raises the exciting possibility of using cosmological observations to constrain not only the absolute mass scale but also the hierarchy (in the case of non-degenerate masses). A number of important questions then arise. Will future or current CMB T+P+WL experiments be able to determine the hierarchy by themselves, or will the inclusion of external data-sets, with their inevitable systematic uncertainties be necessary? What limits the ability of cosmology to determine the hierarchy, and how can these limits be circumvented? Several studies have already attempted to answer some of these questions~\citep{2012arXiv1205.5223O,jimenez_10, debernardis_09, slosar_06, hannestad_03}. Four of these papers chose to parametrize the hierarchy continuously, either by the fraction of the total mass in $m_3$ or using $\Delta = (m_1 - m_3)/\Sigma m_\nu$, with $m_1=m_2$. However, as the constraints in equation~(\ref{eq:masssq}) indicate, the uncertainty on the mass-squared differences is small, and provides good evidence that the hierarchy is \emph{either} normal \emph{or} inverted. In other words, the choice of hierarchy is not a continuous parameter, but a discrete one: with an oscillation prior, the $\Sigma m_\nu$--$\Delta$ space breaks up into two disjoint, one-dimensional spaces and the simplest way to proceed is to analyse any cosmological dataset in both normal \emph{and} inverted models with the masses parametrized by $\sum m_\nu$.


The problem of determining the mass hierarchy is then one of Bayesian model selection, rather than parameter forecasting. The machinery in this case is provided by the Bayes' factor, which quantifies the degree to which different models are favoured by the data with respect to one another, when all their associated parameters are marginalised over. In this work, we investigate the ability of future CMB experiments to determine simultaneously the absolute neutrino mass scale and the (non-parametric) hierarchy. We first calculate forecasts for CMB-only T+P+WL with \textit{COrE} using both Markov-chain Monte Carlo (MCMC) and Fisher techniques, and then include geometric information from future BAO constraints from the full SDSS-III BOSS~\citep{2009astro2010S.314S}, and a Type 1a supernovae survey with \textit{WFIRST}~\citep{2011arXiv1108.1374G}.  We also include a future prior on the Hubble constant. We do not consider information from clustering in this work, such as the matter power spectrum from redshift surveys, number counts, or cosmic shear. Our results are thus immune to the inherent systematic uncertainties in these techniques.

The paper is organised as follows. In Section~\ref{sec:background} we discuss the cosmological influence of both the total neutrino mass and the individual neutrino masses, and the relevant degeneracies with the other cosmological parameters. In Section~\ref{sec:stats} we discuss our statistical forecasting and model selection methodology. In Section~\ref{sec:priors} we introduce the future datasets which will be available as priors for future CMB experiments, and we present and discuss our results in Section~\ref{sec:results}.  Appendices detail our scheme for protecting degeneracies in the construction of the Fisher matrix and discuss the sampling errors in estimates of parameter covariance matrices from MCMC samples.

\section{Cosmological signatures of neutrino masses}
\label{sec:background}

\subsection{Massive neutrinos}

\begin{figure}
\begin{center}
\includegraphics[width=0.7\columnwidth,angle=-90]{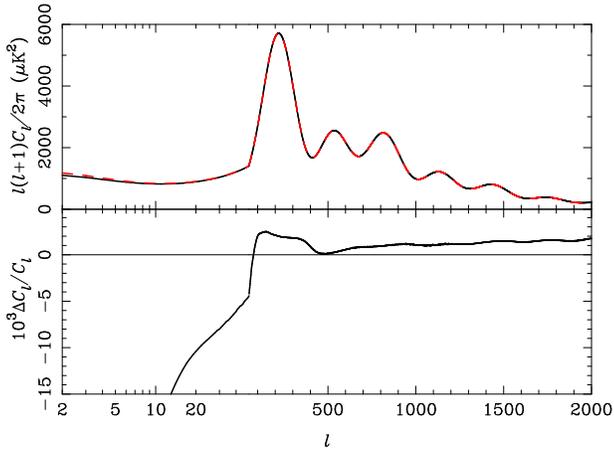}
\end{center}
\caption{\emph{Upper}: Unlensed CMB temperature power spectra for a model with massless neutrinos (dashed red) and degenerate massive neutrinos with $\sum m_\nu = 0.37\,\mathrm{eV}$ (solid black). Both models are flat, have the same physical densities in cold dark matter and baryons, but different Hubble constants ($H_0 = 67.93\,\mathrm{km}\,\mathrm{s}^{-1}\,\mathrm{Mpc}^{-1}$ for the massive case and $H_0 = 71.43\,\mathrm{km}\,\mathrm{s}^{-1}\,\mathrm{Mpc}^{-1}$ for the massless case) to preserve the angular scale of the acoustic peaks. \emph{Lower}: Fractional difference between the massive and massless model. Note that the $x$-axis is logarithmic for $l<50$ and linear for $l\geq 50$.}
\label{fig:Cls_mass_nmass}
\end{figure}

Massive neutrinos have a small, but measurable effect on both the primary anisotropies of the CMB and the growth of structure~\citep[for reviews see][]{elgaroy_05,LesgourguesReview, HannestadReview}. We first consider the primary anisotropies. Since the r.m.s.\ momentum of a neutrino of mass $m_\nu$ at temperature $T_\nu(z)$ satisfies
\begin{equation}
\frac{\langle p_\nu^2 \rangle^{1/2}}{m_\nu} \approx 3.22 \frac{k_B T_\nu(z)}{m_\nu} = \frac{5.5 \times 10^{-4}}{(m_\nu/ \mathrm{eV})}(1+z) \, ,
\end{equation}
neutrinos with mass less than $0.5\,\mathrm{eV}$ are still relativistic at recombination. Their effect on the pre-recombination dynamics of the background and perturbations is thus very similar to the massless case and so they impact the anisotropies only indirectly through the angular diameter distance to last-scattering $d_A(z_*)$. The ratio of $d_A(z_*)$ to the sound horizon at last-scattering $r_s(z_*)$ sets the angular scale of the CMB acoustic peaks. For light masses, the change in sound horizon is small but, if all other physical densities are held fixed, $d_A(z_*)$ falls with increasing mass due to the increased expansion rate at late times. The last-scattering therefore appears closer and the anisotropies are shifted to larger angular scales. In the flat models considered here, this effect is degenerate with a change in either the dark energy density (or, equivalently, the Hubble constant) or the dark energy equation of state parameter $w$ \citep{EB99}. An example of this degeneracy is shown in Fig.~\ref{fig:Cls_mass_nmass} for flat models with $w=-1$ and either massless neutrinos or degenerate massive neutrinos with $\sum m_\nu = 0.37\,\mathrm{eV}$. These models cannot be distinguished on the basis of their (unlensed) spectra. However, since the Hubble constants differ by 4\%, current priors on $H_0$ (e.g. \citealt{2011ApJ...730..119R}) would effectively break the degeneracy between these models.

\begin{figure}
\begin{center}
\includegraphics[width=0.7\columnwidth,angle=-90]{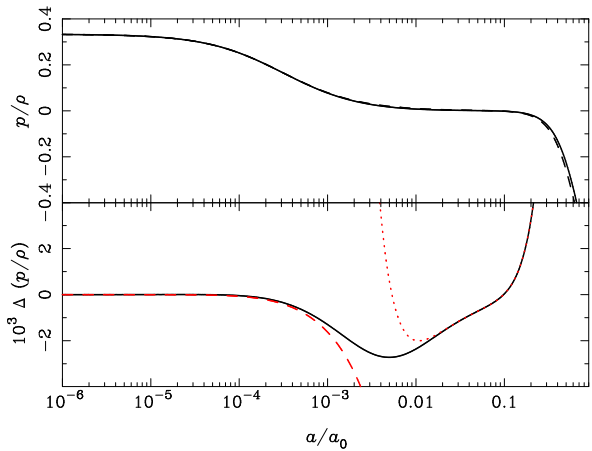}
\end{center}
\caption{\emph{Upper}: Evolution of $p_{\mathrm{tot}}/\rho_{\mathrm{tot}}$ with scale
factor $a$ for the massive (solid) and massless (dashed) models in Fig.~\ref{fig:Cls_mass_nmass}. \emph{Lower}: Difference between $p_{\mathrm{tot}}/\rho_{\mathrm{tot}}$ in these models. Also plotted are the relativistic approximation from equation~(\ref{eq:prho_rel}) and the non-relativistic approximation from equation~(\ref{eq:prho_nonrel}). The differences in $p_{\mathrm{tot}}/\rho_{\mathrm{tot}}$ at late times are due to the reduced energy density in dark energy in the massive model to preserve the angular scale of the CMB acoustic peaks.}
\label{fig:prho}
\end{figure}

The degeneracy is explored in detail in~\cite{2012JCAP...04..027H}. It is not exact due to a number of physical effects, most notably the late-time integrated-Sachs-Wolfe (ISW) effect; see Fig.~\ref{fig:Cls_mass_nmass}. This arises from the late-time decay of the (Weyl) gravitational potential $\phi + \psi$  once dark energy dominates the dynamics of the expansion. On large scales (and for adiabatic initial conditions) the evolution of the gravitational potential $\phi$ follows from constancy of the comoving-gauge curvature perturbation
\begin{equation}
\mathcal{R} = - \phi - \frac{2}{3} \left(\frac{\rho_\mathrm{tot}}{\rho_\mathrm{tot} + p_\mathrm{tot}}\right) \left(\frac{\dot{\phi}}{\mathcal{H}} + \psi\right) \, ,
\end{equation}
where $\rho_\mathrm{tot}$ and $p_\mathrm{tot}$ are the total density and
pressure (including contributions from dark energy), $\mathcal{H}$ is the conformal Hubble parameter and dots denote derivatives with respect to conformal time. The metric potential $\psi = \phi$ at late times when anisotropic stresses can be neglected. The potential is constant for constant
$p_\mathrm{tot}/\rho_\mathrm{tot}$, but evolution in the latter after last-scattering sources the ISW. In Fig.~\ref{fig:prho} we plot $p_\mathrm{tot}/\rho_\mathrm{tot}$ as a function of the scale factor $a$ for the two models considered in Fig.~\ref{fig:Cls_mass_nmass}. The dominant effect in both cases is from the radiation--matter transition and the onset of dark energy domination. The former causes $\phi + \psi$ to decay around last-scattering, sourcing the early-ISW effect which makes a significant contribution to the temperature power spectrum around the first acoustic peak. As dark energy dominates, $p_\mathrm{tot}/\rho_\mathrm{tot} \rightarrow -1$ from nearly zero during the matter-dominated era causing further decay of the potentials and sourcing the late-time ISW effect.
Massive neutrinos change the picture in the following ways. Compared to massless neutrinos, $\rho_\nu$ increases as they become non-relativistic and $p_\nu$ decreases; the onset of this transition is described by~\citep{lewischallinor_02}
\begin{eqnarray}
\rho_\nu &\approx & \rho_\nu^0 \left[1+ \frac{5}{7\pi^2}\left(\frac{m_\nu}{k_B T_\nu(z)}\right)^2 \right]\, , \nonumber \\
p_\nu &\approx & \frac{\rho_\nu^0}{3} \left[1- \frac{5}{7\pi^2}\left(\frac{m_\nu}{k_B T_\nu(z)}\right)^2 \right]\, ,
\label{eq:prho_rel}
\end{eqnarray}
where $\rho_\nu^0$ is the energy density per species of massless neutrino.
In the non-relativistic limit~\citep{lewischallinor_02},
\begin{eqnarray}
\rho_\nu &\approx & \frac{180\rho_\nu^0}{7\pi^4} \left[\zeta(3)\left(\frac{m_\nu}{k_B T_\nu(z)}\right) + O\left(\frac{k_B T_\nu(z)}{m_\nu}\right) \right]\, , \nonumber \\
p_\nu &\approx & \frac{900\rho_\nu^0}{7\pi^4} \left[\zeta(5)\left(\frac{k_B T_\nu(z)}{m_\nu}\right) + O\left(\frac{k_B T_\nu(z)}{m_\nu}\right)^3 \right]\, .
\label{eq:prho_nonrel}
\end{eqnarray}
The ratio $p_{\mathrm{tot}}/\rho_{\mathrm{tot}}$ initially falls more quickly
in models with massive neutrinos and this leads to enhanced decay of the Weyl
potential around last-scattering and a larger early ISW effect (see Fig.~\ref{fig:Cls_mass_nmass}). If the angular scale of the acoustic peaks and the physical densities in cold dark matter and baryons are fixed (the latter preserving the pre-recombination physics), the fraction of energy density in dark energy is less in models with massive neutrinos and at late times $p_{\mathrm{tot}}/\rho_{\mathrm{tot}}$ falls less slowly towards $-1$. This reduces the late ISW effect.

The direct effect of neutrino masses on the pre-recombination physics is very small for $m_\nu \ll 0.5\,\mathrm{eV}$. The enhanced energy density reduces the sound horizon and damping scale, with the former being compensated by changes in $d_A(z_*)$ if we fix the angular scale of the acoustic peaks. The neutrino perturbations themselves influence the CMB via the back-reaction from their stress-energy on the metric perturbations. The corrections are $O(1-v_\nu)$ where $v_\nu = 1 - O(m_\nu/k_B T_\nu)^2$ is the typical neutrino thermal velocity. The size of the effects in the neutrino stress-energy tensor is $O(m_\nu / k_B T_\nu)^2
\rho_\nu^0$ and the relative importance for the metric perturbations and the CMB is $O(m_\nu / k_B T_\nu)^2\rho_\nu^0/\rho_\mathrm{tot}$. For our model with $\sum m_\nu = 0.37\, \mathrm{eV}$, we expect effects at the $0.1\%$ level consistent with the small residual differences at high $l$ in the CMB power spectrum shown in Fig.~\ref{fig:Cls_mass_nmass}.

We end our discussion of the primary anisotropies by noting that the effects of 1--$10\,\mathrm{eV}$ masses on the CMB are rather different, since the neutrinos are already non-relativistic at the time of recombination~\citep{dodelsongatesstebbins}. At these masses, which are already ruled out by current upper limits on the total mass, the neutrinos act as a hot dark matter component.

The matter power spectrum is also affected by the presence of massive
neutrinos~\citep{1980PhRvL..45.1980B}. Once non-relativistic, they increase the expansion rate over the massless case, but the tendency of this to impede growth in the clustering of the other matter components is mitigated on scales where the neutrinos can cluster. At any time, neutrinos can cluster on scales larger than their proper Jeans (or free-streaming) length, approximately $v_\nu(z) / H(z)$. For massless neutrinos, this is simply the particle horizon and the comoving Jeans length grows in time in a non-accelerating universe. However, for non-relativistic neutrinos, the free-streaming length~\citep{LesgourguesReview}:
\begin{equation}
\lambda_{\mathrm{FS}}(z) \approx 7.7 \frac{1+z}{\sqrt{\Omega_{\Lambda}+\Omega_{\mathrm{m}}(1+z)^3}} \left(\frac{1\mbox{ eV}}{m_\nu} \right) h^{-1}\, \mbox{Mpc}\, ,
\label{eq:lambdafs}
\end{equation}
where $\Omega_m$ is the current density parameter for matter including non-relativistic massive neutrinos. For a non-relativistic transition in matter domination, $\lambda_{\mathrm{FS}}(z) \sim a^{1/2}$, so the comoving free-streaming length \emph{decreases} in time. Therefore, the comoving free-streaming wavenumber [$k_{\mathrm{FS}} \equiv 2\pi a/\lambda_{\mathrm{FS}}(z)$] has a minimum given by the comoving scale of the horizon at the non-relativistic transition. On scales larger than this, neutrinos have always clustered and their mass has no effect on the matter power spectrum. Below the comoving horizon at the non-relativistic transition, neutrinos only cluster only after they exit the (shrinking) comoving free-streaming length thus slowing down the growth of structure in the intervening time. On all scales smaller than the current free-streaming scale, massive neutrinos are not clustered at the present time giving a scale-free fractional suppression of the matter power power spectrum by roughly $-8f_\nu$, where
$f_\nu = \Omega_\nu / \Omega_m$~\citep{1998PhRvL..80.5255H}.

For sub-eV neutrino masses, $k_{\mathrm{FS}}$ today is at roughly the same scale that non-linear corrections to the matter power spectrum begin to become important~\citep{HannestadReview}. $N$-body simulations indicate a larger suppression of around $-9.6f_{\nu}$ at scales $k \approx 0.5$--$ 1 h\, \mathrm{Mpc}^{-1}$, accurate to about 1\%~\citep{brandbyge_and_hannestad_09, viel_et_al_10}. Note that this is considerably smaller than the BAO scale ($k \approx 0.15h\, \mathrm{Mpc}^{-1}$).

\begin{figure}
  \centering
  \includegraphics[width=0.5\textwidth]{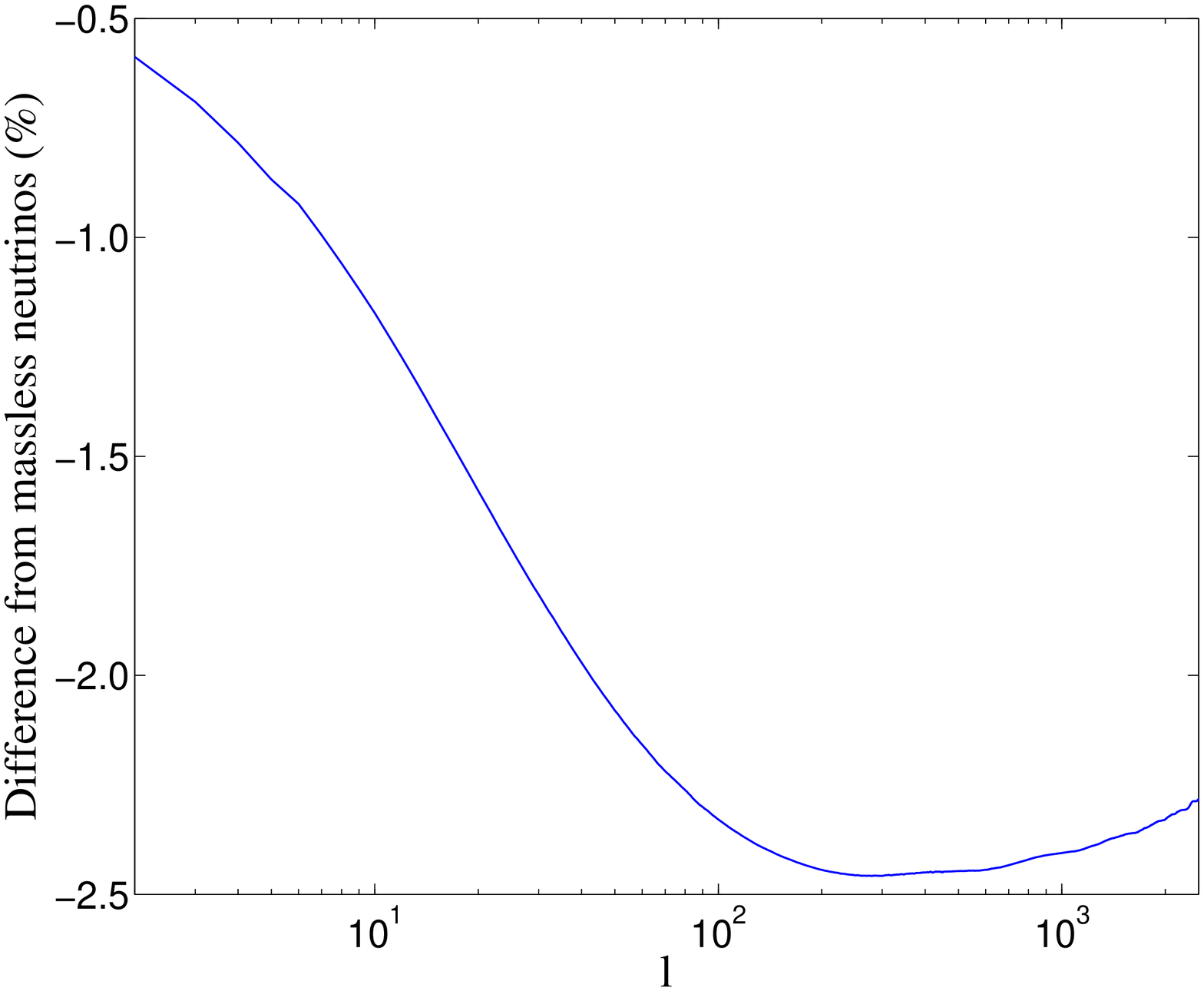}
  \includegraphics[width=0.5\textwidth]{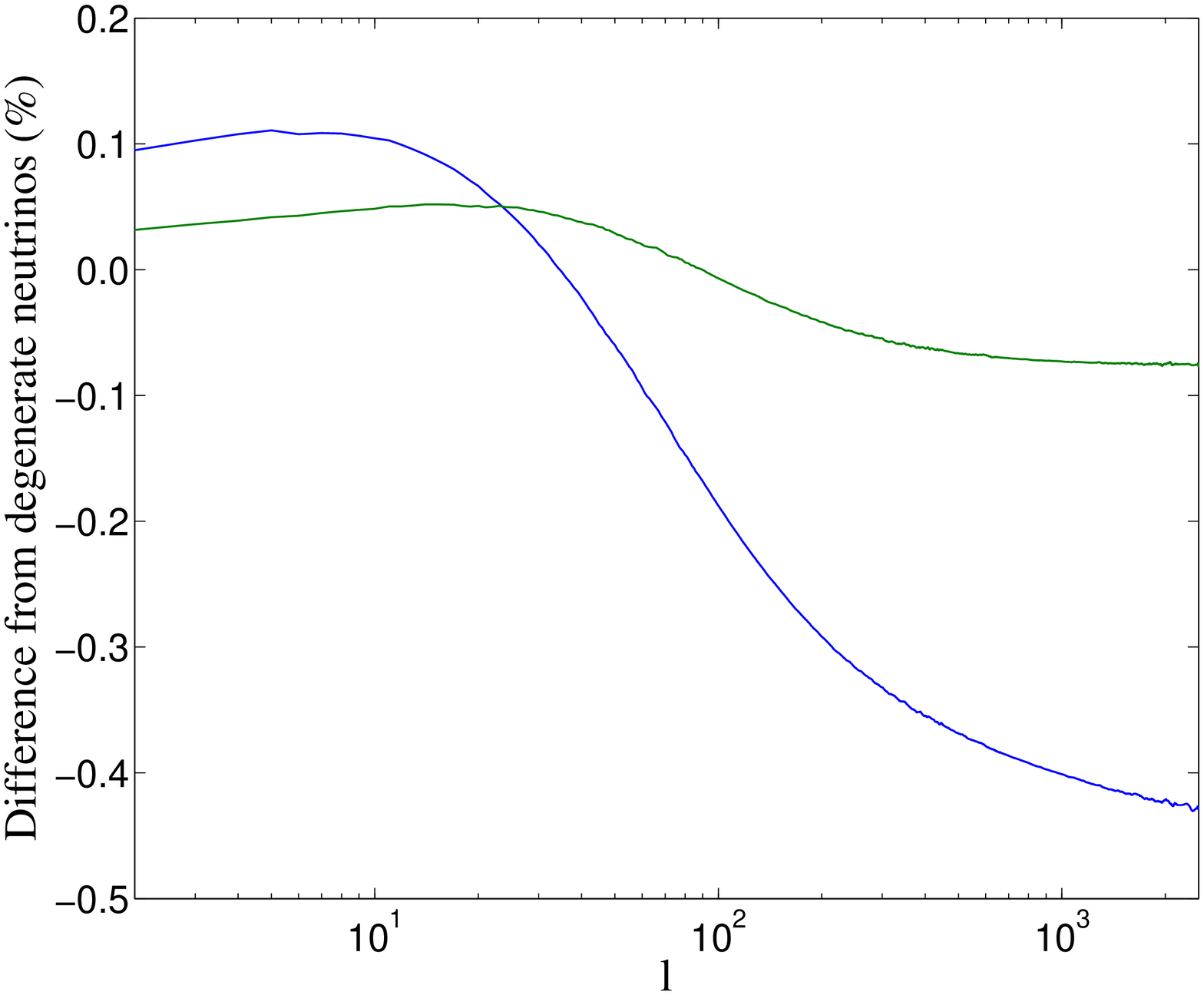}
  \caption{\textit{Upper}: Fractional difference of the lensing power spectrum from a scenario with massless neutrinos, for a total mass of $0.095\,\mathrm{eV}$. The suppression of power on small scales is clearly seen. 
\textit{Lower}: Fractional difference of the lensing power spectrum from a scenario with degenerate neutrinos, for a fixed total mass of 0.095 eV in the inverted hierarchy (blue) and normal hierarchy (green).}
  \label{fig:lensing}
\end{figure}

One of the key observables considered in this paper is the reconstruction of the CMB weak-lensing deflection field from the lensed primary anisotropies \citep[for a review, see][]{anthonyreview}. In linear theory, the power spectrum of the lensing deflection angle, $C_l^{dd}$, is a line-of-sight integral over the matter power spectrum, and so the effect of massive neutrinos is similar to their effect on the growth of large-scale structure.
The suppression of small-scale power is still the primary effect, as shown in the upper plot of Fig.~\ref{fig:lensing}, where we plot the fractional change in the lensing power spectrum in the presence of massive neutrinos compared to the massless case. The total mass is $0.095\,\mathrm{eV}$,
the minimum mass of the inverted hierarchy. The differences are at the few percent level, significantly larger than the intrinsic effect in the unlensed CMB anisotropies which is at the $0.01\%$ level (for $l \geq 100$) for such masses. Moreover, the information is complementary since the effect in the lensing power spectrum scales roughly with the total mass while the intrinsic effect in the unlensed CMB scales with the square of the masses.
Neutrinos also have a small effect on the cross-correlation of the lensing deflection with the CMB temperature anisotropy (due to the late-time ISW effect) and the CMB $E$-mode polarisation~\citep{lewischallinorhanson_11}, and we include both in our analysis for completeness.

%

\subsection{Individual masses}

As we have seen, sub-eV massive neutrinos mostly affect the primary CMB anisotropies through their effect on the angular-diameter distance $d_A(z_*)$. The masses are therefore degenerate with other late-time parameters (such as the dark energy model). Even if the other late-time parameters are fixed by external distance data, there is very little sensitivity to individual masses since $d_A(z_*)$ is determined primarily by the summed mass $\sum m_v \approx 93.14 \Omega_\nu h^2\,\mathrm{eV}$.

The signature of mass \emph{differences} amongst neutrinos is thus felt mainly through their effect on the growth of structure. Different masses have different free-streaming wavenumbers, and each has their own unique signature on the structure formation history of the universe. As extremes, if all the mass were in one eigenstate the non-relativistic transition would be earlier than if the masses were degenerate and the damping in the matter power spectrum would not extend to such large scales. On scales smaller than, but close to, the horizon size at the non-relativistic transition for the degenerate case, the matter power spectrum should therefore be smaller for the degenerate case compared to if all mass were in one eigenstate. However, this behaviour reverses on scales below the smaller (i.e.\ the non-degenerate case) of the free-streaming scales at the observed redshift, and the degenerate case would have less suppression of power. This is because on such scales the neutrinos would never have clustered since early times and the degenerate case would have a later non-relativistic transition and a slightly lower neutrino energy density through the extended transition epoch (see Figs~3 and~4 in~\citealt{2004PhRvD..70d5016L}). These signatures also show up in the CMB WL power spectrum, which roughly reflects the matter power spectrum around $z \sim 2$. This is illustrated in the lower plot of Fig.~\ref{fig:lensing}, where we plot the fractional difference in the deflection power spectrum compared to the case of degenerate masses in the two hierarchies at fixed total mass $\sum m_\nu = 0.095\,\mathrm{eV}$. The situation is rather more subtle than the extreme cases discussed above, since there are effectively two free-streaming scales and non-relativistic transitions in the normal hierarchy at this total mass. Although cosmic variance (3\% at $l = 1000$) dominates the differences in deflection power between the two hierarchies at each multipole, the broad-band nature of the signal means that we can combine many multipoles to beat down cosmic variance (to roughly $0.1\%$ for all multipoles up to $l=1000$).

It should therefore be possible, at least in principle, to determine the mass hierarchy from a combination of CMB T+P and CMB WL observations. Including the oscillation measurements as prior information significantly ameliorates this task, as we only have to determine a single model from a choice of two, rather than deduce the mass splittings purely from cosmology. However, as we shall see, parameter degeneracies present a considerable obstacle to realising this goal.

Finally, we note that it has recently been claimed~\citep{2012arXiv1203.5342W} that non-linearities roughly double the matter power spectrum differences between the hierarchies for a fixed total mass appropriate to the mininal-mass inverted hierarchy ($\sum m_\nu = 0.095\,\mathrm{eV}$) on mildly non-linear scales ($k \sim 1 h\,\mathrm{Mpc}^{-1}$). However, such small scales are deep in the region where CMB lensing reconstructions will always be dominated by statistical noise; see Section~\ref{sec:fisher}.

\section{Statistical Methods}
\label{sec:stats}

\subsection{Markov-chain Monte Carlo}

The principal tool we use to forecast parameter constraints is Markov-chain Monte Carlo (MCMC). We first generate fiducial unlensed CMB spectra, $C_l^{TT}$, $C_l^{TE}$ and $C_l^{EE}$, the lensing deflection power spectrum, $C_l^{dd}$, and the cross-correlations, $C_l^{Td}$ and $C_l^{Ed}$, using the publicly-available Boltzmann code \textsc{CAMB}~\citep{camb} with fiducial parameters $\theta_0$. We then analyse the mean log-likelihood as a function of parameters $\theta$, given by
\begin{equation}
-2\langle\ln{P(\theta|\theta_0)}\rangle = \rmn{Tr}[\mathbfss{C}(\theta_0)\mathbfss{C}^{-1}(\theta)] + \ln \frac{|\mathbfss{C}(\theta)|}{|\mathbfss{C}(\theta_0)|} - \rmn{dim}(\mathbfss{C}),
\label{eq:log_like}
\end{equation}
where $\mathbfss{C}$ is the covariance matrix of the data vector $\bmath{d}=[a_{lm}^T,a_{lm}^E,a_{lm}^d]$, which consists of the unlensed temperature and $E$-mode polarization, and the reconstructed lensing deflection multipoles. Note that $\mathbfss{C}$ includes instrument noise and the statistical noise of the lensing reconstruction. The mean log-likelihood has been normalised to zero at the maximum-likelihood point, where $\theta=\theta_0$. Such a likelihood was considered in e.g.~\citet{lindsay} in the context of extracting cluster masses from CMB lensing.

\begin{table}
  \caption{Fiducial parameters used throughout this work. Note that the density parameters for massive neutrinos are close to the values in the minimal-mass normal and inverted hierarchies: they correspond to total masses $\sum m_\nu = 0.062 \, \mathrm{eV}$ (normal) and $\sum m_\nu = 0.105\,\mathrm{eV}$ (inverted).}
  \begin{center}
    \begin{tabular}{c c}
      \hline
      Parameter & Value \\
      \hline
      $\Omega_bh^2$ & 0.023 \\
      $\Omega_{c}h^2$ & 0.112 \\
      $h$ & 0.703 \\
      $\tau$ & 0.085 \\
      $A_s$ & 2.42$\times10^{-9}$ \\
      $n_s$ & 0.966 \\
      $\Omega_{\nu}h^2$ (normal) & 0.000666 \\
      $\Omega_{\nu}h^2$ (inverted) & 0.00113 \\
      $w$ & -1\\
      $100\theta(z_*)$ (normal) & 1.03957 \\
      $100\theta(z_*)$ (inverted) & 1.04081 \\
      \hline
      \end{tabular}
    \end{center}
  \label{table:fid_model}
\end{table}

\begin{table}
  \caption{Central frequencies $\nu$, beam full-widths at half-maximum $\theta_{\mathrm{fwhm}}$, and temperature and polarization sensitivities, $\Delta_T$ and
$\Delta_P$, for the CMB channels of \textit{COrE}~\citep{core}.}
  \begin{center}
    \begin{tabular}{c c c c}
      \hline
      $\nu$ & $\theta_{\mathrm{fwhm}}$ & $\Delta_T$ & $\Delta_P$ \\
      (GHz) & (arcmin) & ($\mu\mathrm{K}\,\mathrm{arcmin}$) & ($\mu\mathrm{K}\,\mathrm{arcmin}$) \\ 
      \hline
      105 & 10.0 & 2.7 & 4.6 \\
      135 & 7.8 & 2.6 & 4.5 \\
      165 & 6.4 & 2.6 & 4.6 \\
      195 & 5.4 & 2.6 & 4.5 \\
      225 & 4.7 & 2.6 & 4.5 \\
      \hline
      \end{tabular}
  \end{center}
  \label{table:noise}
\end{table}
\begin{figure}
\centering
\includegraphics[width=0.35\textwidth,angle=-90]{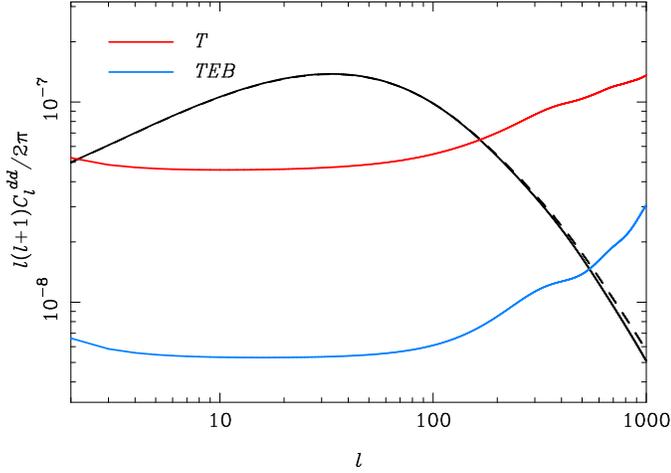}
\caption{Power spectrum of the statistical noise on lensing reconstructions for \textit{COrE} using only temperature information (red) or temperature and polarization (blue). The linear-theory lensing deflection power spectrum is also shown (black), along with the effect of including the non-linear matter power spectrum (black dashed).}
\label{fig:lensing_noise}
\end{figure}

Our fiducial parameters are given in Table~\ref{table:fid_model}, and are the maximum likelihood estimates from the WMAP 7-year release~\citep{wmap7} except for the density parameters for massive neutrinos which we take to be close to the values for the minimal-mass normal and inverted hierarchies. We use noise levels appropriate to the proposed \textit{COrE} mission~\citep{core}; see Table~\ref{table:noise}. We account crudely for removal of astrophysical foregrounds by using only the frequency channels in the range 105--$225\,\mathrm{GHz}$ and retaining a fraction $f_{\mathrm{sky}}=0.7$ of the sky. We implement the latter approximately by reducing the number of degrees of freedom per observable at each multipole from $2l+1$ to $(2l+1)f_{\mathrm{sky}}$ when evaluating equation~(\ref{eq:log_like}). We retain multipoles up to $l_{\mathrm{max}}=2500$.
To compute the statistical noise on the lensing deflection reconstruction, we use the optimal quadratic estimator of~\citet{okamotoandhu}; the noise power spectrum is shown in Fig.~\ref{fig:lensing_noise} for reconstructions from~\textit{COrE} using only temperature or temperature and $E$ and $B$-mode polarization. For \textit{COrE}, the $E$-$B$ estimator dominates the reconstruction.

Note that we use the \emph{unlensed} temperature and $E$-mode polarization in the likelihood. These are not directly observable but we use them as a simple work-around of the double-counting issues that may arise when jointly analysing the lensed CMB and lens reconstruction.

We impose the prior $w \ge -1$ on the dark energy equation of state. This represents prior preference for a model in which dark energy is described by a field satisfying the weak energy condition. Models with $w < -1$ do exist, but possess greater complexity~\citep{PhysRevD.78.087303}, and are potentially unstable at the quantum level~\citep{carroll03}. Since the data do not prefer $w$ either side of $-1$, we select the simpler model, $w \ge -1$.

We sample from the likelihood using the publicly-available \textsc{CosmoMC} package~\citep{cosmomc}, modified to include the effects of neutrino mass splitting. We use a modified version of the \textsc{FUTURCMB} lensing add-on for \textsc{CosmoMC}~\citep{perotto_06} including the small correlation between the $E$-mode polarization and the lensing deflection~\citep{lewischallinorhanson_11} calculated in \textsc{CAMB} as $C_l^{Ed}$. \textsc{CosmoMC} works natively with the angular scale of the sound horizon at recombination, $\theta(z_*)$ as opposed to $h$, so we include the fiducial value of this parameter in Table~\ref{table:fid_model} for completeness.

When forecasting standard deviations of parameters, we must 
estimate the covariance matrix from the MCMC chains.  For those 
parameters whose marginalised posteriors are approximately 
Gaussian, this is the quantity that should be compared to the 
results of a Fisher analysis (parameters with hard priors do not 
fall into this category and so do not require standard deviation 
estimates). However, these estimates are inherently statistical in nature.
If we wish to compare the results of the MCMC analysis to those of a Fisher
analysis we need to know the typical
statistical fluctuation of our covariance matrix estimates. This 
translates into some scatter in the eigenvalues and eigenvectors, a matter
that we explore in Appendix~\ref{app:covmat}.

\subsection{Fisher analysis}
\label{sec:fisher}

In Section~\ref{sec:results}, we make some comparisons between parameter constraints obtained with MCMC and those from a Fisher analysis. The Fisher information matrix is the Hessian of (minus) the mean log-likelihood at the fiducial parameters. To the extent that the likelihood is Gaussian in the parameters, constraints derived from MCMC exploration of the mean log-likelihood should agree with the Fisher analysis. Differentiating equation~(\ref{eq:log_like}) with respect to the parameters gives the Fisher matrix~\citep[see, for example][]{moderncosmology}
\begin{eqnarray}
F_{ij} &=& \frac{1}{2}\rmn{Tr}\left[\mathbfss{C}^{-1}
\frac{\partial \mathbfss{C}}{\partial \theta_i}
\mathbfss{C}^{-1}
\frac{\partial \mathbfss{C}}{\partial \theta_j} \right] \nonumber \\
&=& \sum_{l=2}^{l_{\mathrm{max}}}\sum_{XX',YY'}\frac{\partial \mbox{C}_l^{XX'}}{\partial \theta_i}[\mbox{Cov}(XX',YY')]^{-1}_l\frac{\partial \mbox{C}_l^{YY'}}{\partial \theta_j}, 
\label{eq:fisher_matrix}
\end{eqnarray}
where $\mbox{Cov}(XX',YY')$ is the covariance of the power spectra estimators, including noise, and $XX'$ and $YY'$ stand for the spectra $TT$, $EE$, $TE$, $dd$, $Td$, and $Ed$, with $dd$ the weak lensing power spectrum etc. Assuming Gaussian fields and noise, the specific form of $\mbox{Cov}(XX',YY')$ is
\begin{equation}
[\rmn{Cov}(XX',YY')]_l=\frac{1}{(2l+1)f_{\rmn{sky}}}\left(\mbox{\~{C}}_l^{XY}\mbox{\~{C}}_l^{X'Y'}+\mbox{\~{C}}_l^{XY'}\mbox{\~{C}}_l^{X'Y}\right),
\label{eq:paramcov}
\end{equation}
where $f_{\mathrm{sky}}$ is the fractional sky coverage. The tildes denote
the total power spectra including instrument noise for $XX'=TT$ and $EE$
and reconstruction noise for $XX'=dd$.
The inverse of the Fisher matrix gives the covariance matrix between the parameters and its diagonal elements give the $1\sigma$ marginalised errors on parameters.

When constructing the Fisher matrix, it is important to use accurate power spectrum derivatives, since numerical noise in these can artificially break degeneracies leading to over-optimistic parameter constraints. We found this particularly troublesome for the lensing deflection field derivatives. Our brute-force solution was to run \textsc{CAMB} at a high accuracy setting (\texttt{accuracy\_boost}=5) to remove this noise, as it was found not to be due to a bad choice of derivative step size\footnote{We used a pre-January 2012 version of \textsc{CAMB} for this work. Several improvements in numerical accuracy were made for the January 2012 version; see~\cite{2012JCAP...04..027H}.}. In addition, we found it necessary to enforce some parameter degeneracies directly in the construction of the Fisher matrix. We detail these issues in Appendix~\ref{app:fisher}.

\subsection{Bayesian model selection}
\label{sec:modelselection}

Distinguishing between the two hierarchies is a problem of model selection. We can quantify our relative degree of belief in different models by use of the Bayes' factor, defined as the ratio of Bayesian evidences. Consider two models $M$ and $M'$ (for example, normal and inverted hierarchies), with parameter vectors $\bmath{\theta}$ and $\bmath{\theta}'$, not necessarily of the same dimension. Let $\bmath{x}$ be the data vector. The ratio of posterior probabilities is
\begin{equation}
\frac{p(M'|\bmath{x})}{p(M|\bmath{x})} = 
\frac{p(\bmath{x}|M')p(M')}{p(\bmath{x}|M)p(M)},
\end{equation}
where $p(M)$ is our prior degree of belief in the model $M$ and similarly for $M'$. The evidence $p(\bmath{x}|M)$ is found by marginalising the product of the likelihood $p(\bmath{x}|\bmath{\theta},M)$ and the parameter prior $p(\bmath{\theta}|M)$ over parameters. Assuming no \textit{a priori} preference for either model, $p(M)=p(M')$, and so the ratio of posterior probabilities becomes the Bayes' factor
\begin{equation}
B \equiv \frac{\int d \bmath{\theta}' p(\bmath{x}|\bmath{\theta}',M')
p(\bmath{\theta}'|M')}{\int d \bmath{\theta} p(\bmath{x}|\bmath{\theta},M)
p(\bmath{\theta}|M)} \, .
\end{equation}
Evaluating $B$ involves a difficult multi-parameter integration but we may make some analytical progress by approximating the likelihood as Gaussian, the `Laplace approximation' \citep[for example, ][]{mackay}. For the case of $\sum m_{\nu}$ and $w$, the prior cuts off the posterior at close to the maximum likelihood point, so the integration over these parameters must be performed numerically. Starting from our MCMC chain samples, we perform a least-squares fit of a multi-variate Gaussian to the likelihood. For model selection, we parametrize the neutrino masses in terms of the lightest mass, rather than the total mass, so that the prior volumes for the two hierarchies are equal. We calculate the evidence assuming a uniform prior in the lightest mass (and the other parameters) by first integrating analytically the Gaussian fit over all parameters except $w$ and the mass. The resulting two-dimensional Gaussian is integrated over the prior range numerically. The Bayes' factor then reduces to a product of the maximum likelihood ratio and an `Occam factor' in analogy to the fully Gaussian case. However, here the Occam factor derives from both the Gaussian covariance and the degree to which the likelihood is cut off by the prior.

We perform model comparison in the case of the minimal-mass inverted hierarchy taking the `data' to be the fiducial power spectra (plus noise) calculated in the inverted hierarchy. When analysed assuming the normal hierarchy, the likelihood has a local maximum within the prior volume. We locate this point, and the likelihood there, leaving us to fit only the covariance of the multi-variate Gaussian. The Bayes' factor we calculate is the ratio of the evidences derived from the respective mean log-likelihoods. Although this will generally differ from the mean Bayes' factor, we denote it by $\langle B \rangle$ noting that its value should be typical in an ensemble of data from the minimal-mass inverted hierarchy.



The smaller the volume of the likelihood confidence ellipsoid, the more finely-tuned the model must be to fit the data. Such models are penalised in the Bayes' factor in favour of models which do not need such fine-tuning~\citep{mackay}. The other term in the Bayes' factor, involving the ratio of the maximum likelihoods, represents the data's sensitivity to the mass splitting. If cosmology were insensitive to mass splittings, the ratio of maximum likelihoods would always be unity if the true total mass were large enough to be realised in either hierarchy. In this case, the Occam factor will generally favour the normal hierarchy as the likelihood is non-zero over a greater range of the prior volume.


\citet{jeffreys} proposed model selection criteria depending on the value taken by the Bayes' factor. If $\ln B > 5$, evidence for model $M'$ is `decisive' over model $M$, if $2.5 < \ln B < 5$ the evidence is `strong', and if $1 < \ln B < 2.5$ it is `substantial'. For a discussion of the suitability of these criteria, and the usefulness of $B$ as a statistic, see \citet{efstathiou_08} and \citet{jenkinsandpeacock}.
\section{External Datasets}
\label{sec:priors}

In this section we discuss the various priors from non-CMB data that we include in our analysis. For a comprehensive survey of the utility of external data in constraining neutrino masses, see~\citet{abazajianreview}. We consider only a subset of all possible probes, since the main focus of this paper is information from primary CMB anisotropies with a weak lensing reconstruction. CMB experiments offer a relatively clean source of cosmological information at multipoles $l<2000$ since the relevant physics is simple and well-understood and extra-Galactic foregrounds are sub-dominant to the primary CMB fluctuations.

We only include external geometric probes ($H_0$, luminosity distances from supernovae and BAO) here. Other direct probes of the clustering of matter on small scales, such as galaxy clustering, galaxy weak lensing, the Ly-$\alpha$ forest and cluster abundances are potentially very useful probes of neutrino masses because of the scale-dependent growth associated with neutrino free-streaming (see Section~\ref{sec:background}). However, associated problems such as redshift-space distortions, scale-dependent galaxy bias, source redshift uncertainties, and the fact that the free-streaming scale lies close to the non-linearity scale at low redshift make it interesting to see what can be achieved with only the CMB and relatively clean geometric probes.


%
\subsection{Hubble constant}

As we discuss later in Section~\ref{sec:results}, our Fisher matrix indicates that the geometric degeneracy between $h$, $w$ and neutrino mass is not completely broken by lensing. The neutrino mass contributes to this degeneracy in a small way, so we might hope to improve our forecasts with a simple prior on $H_0$. Current precision from the Hubble Space Telescope is around the 3\% level \citep{2011ApJ...730..119R}, but for the purpose of forecasting we impose a 2\% prior, not unrealistic on the time-scale of a mission like \textit{COrE}~\citep{freedman}.
\subsection{\textit{WFIRST} Type 1a supernovae}

Distance modulus information from Type 1a supernovae offers geometric information about the universe which can be a useful probe of late-time phenomena such as dark energy \citep{riess_98, perlmutter_99}, as well as a tool for breaking geometric degeneracies inherent in the CMB \citep{efstathiou_etal_99, lineweaver_98}. 

In this work, we forecast distance modulus measurements from the Wide-Field InfraRed Survey Telescope (\textit{WFIRST}; \citealt{2011arXiv1108.1374G}), the highest ranked recommendation for large space-based missions in the 2010 US Decadal Survey~\citep{national2010New}. Expected survey characteristics were taken from the \textit{WFIRST} Interim Report\footnote{\url{http://wfirst.gsfc.nasa.gov/science/WFIRST_Interim_Report.pdf}}. We assume their `conservative' figure of merit assumption, but double the survey time to 12 months. We forecast 200 supernovae in each of eight redshift bins between $z=0.4$ and $z=1.2$, each bin having $\Delta z=0.1$. We augment this sample with 500 nearby ($z<0.1$) supernovae, as forecast by the Figure of Merit Science Working Group Panel\footnote{\url{http://wfirst.gsfc.nasa.gov/science/fomswg/fomswg_technical.pdf}}. We assume a scatter $\sigma_M(z)=0.11+0.033z$ of the apparent magnitudes for each supernova, after light-curve fitting, about the (unknown) absolute magnitude $M$. The mean absolute magnitude in each redshift bin then has variance $\sigma_{\mathrm{tot}}^2(z_i) = \sigma_M^2(z_i)/N_i + \sigma^2_{\mathrm{sys}}(z_i)$, where $N_i$ is the number of supernovae in each bin and $\sigma_{\mathrm{sys}}^2(z)=0.02(1+z)/1.8$ represents a floor in the scatter due to systematic effects. Our treatment is consistent with the \textit{WFIRST} `conservative' forecasts. We neglect potential biases through use of different light curve fitters, and assume all such uncertainty is contained in our systematic error.

We construct the mean log-likelihood after marginalising over the absolute magnitude $M$ (see Appendix F of~\citealt{cosmomc}). Assuming no correlation between redshift bins, we have, up to a constant,
\begin{equation}
-2 \langle \ln P(\theta|\theta_0)\rangle = \sum_i \frac{\Delta \mu_i^2}{\sigma^2_{\mathrm{tot}}(z_i)} - \left(\sum_i \frac{1}{\sigma^2_{\mathrm{tot}}(z_i)}\right)^{-1} \left(\sum_i \frac{\Delta \mu_i}{\sigma^2_{\mathrm{tot}}(z_i)}\right)^2 \, ,
\label{eq:snloglike}
\end{equation}
where $\Delta \mu_i \equiv \mu(z_i;\theta) - \mu(z_i;\theta_0)$ with $\mu(z_i;\theta)$ the distance modulus at parameters $\theta$, and $\theta_0$ is the fiducial model. The sums are over redshift bins. The second term in equation~(\ref{eq:snloglike}) arises from the marginalisation and ensures that the log-likelihood does not change under $\Delta \mu_i \rightarrow
\Delta \mu_i + M$. 

\subsection{BOSS baryon acoustic oscillations}

The sound horizon at the baryon drag epoch, when baryons were effectively released from photons, imprints a characteristic scale in the matter distribution. Observing projections of this standard ruler in the galaxy distribution allows one to map out $H(z)$ and $d_A(z)$ at a range of redshifts~\citep{blakeandglazebrook, seoandeisenstein_03,huandhaiman}. Current measurements of BAO are limited to the spherically-averaged correlation function (or power spectrum) which is sensitive to an effective distance $[d_A^2(z)/H(z)]^{1/3}$; see, for example,~\citet{2011MNRAS.416.3017B,2011MNRAS.418.1707B,2012arXiv1203.6594A} for the most recent measurements.

A major advance in BAO detection will come from the completion of BOSS~\citep{2009astro2010S.314S}, part of SDSS-III. This should allow separate measurements of the angular diameter distance and Hubble parameter in several redshift bins. In Table~\ref{table:boss} we show the forecast constraints on $d_A(z)$ and $H(z)$, taken from the SDSS-III Project Description\footnote{\url{http://www.sdss3.org/collaboration/description.pdf}}. What is actually measured is $d_A(z)/r_s$ and $H(z)r_s$, where $r_s$ is the sound horizon at the baryon drag epoch, and it is this quantity we compute in our MCMC analysis. The two are mildly correlated, and we assume a correlation coefficient of 0.4, consistent with the value found in \citet{seoandeisenstein_07}.
\begin{table}
\caption{Forecast BOSS errors on $d_A(z)$ and $H(z)$ in three redshifts bins.
}
  \begin{center}
    \begin{tabular}{|c|c|c|}
      \hline
      $z$ & $d_A(z)$ & $H(z)$ \\ \hline
      0.35 & 1.0\% & 1.8\% \\
      0.60 & 1.1\% & 1.7\% \\
      2.50 & 1.5\% & 1.5\% \\ \hline
    \end{tabular}
  \end{center}
  \label{table:boss}
\end{table}
\section{Results}
\label{sec:results}
\subsection{Neutrino mass forecasts}

\begin{table}
  \caption{Parameter errors ($1\sigma$) in the two hierarchies, comparing MCMC and Fisher matrix results. We assume noise levels appropriate to \textit{COrE} and use only CMB T+P+WL. We quote 68\% upper limits on $\sum m_{\nu}$ and $w$.}
  \begin{center}
    \begin{tabular}{c c c c c}
      \hline
       & \multicolumn{2}{c}{Inverted} & \multicolumn{2}{c}{Normal} \\
      \hline
      & MCMC & Fisher & MCMC & Fisher \\
      \hline
      $\Omega_bh^2$ & $3.71\times10^{-5}$ & $3.84\times10^{-5}$ & $3.62\times10^{-5}$ & $3.82\times10^{-5}$\\
      $\Omega_{c}h^2$ & $4.34\times10^{-4}$ & $5.60\times10^{-4}$ & $4.30\times10^{-4}$ & $5.50\times10^{-4}$\\
      $h$ & $0.014$ & $0.041$ & $0.015$ & $0.042$ \\
      $\tau$ & $2.32\times10^{-3}$ & $2.33\times10^{-3}$ & $2.34\times10^{-3}$ & $2.43\times10^{-3}$ \\
      $\log{10^{10}A_s}$ & $7.00\times10^{-3}$ & $8.26\times10^{-3}$ & $6.83\times10^{-3}$ & $8.17\times10^{-3}$ \\
      $n_s$ & $1.63\times10^{-3}$ & $1.86\times10^{-3}$ & $1.62\times10^{-3}$ & $1.86\times10^{-3}$\\
      $\sum m_{\nu} (\mathrm{eV})$ & $<0.136$ & $<0.171$ & $<0.098$ & $<0.151$\\
      $w$ & $<-0.93$ & $<-0.87$ & $<-0.93$ & $<-0.87$\\
      \hline
    \end{tabular}
  \end{center}
  \label{table:errors}
\end{table}

\begin{figure}
\centering
\includegraphics[width=0.5\textwidth]{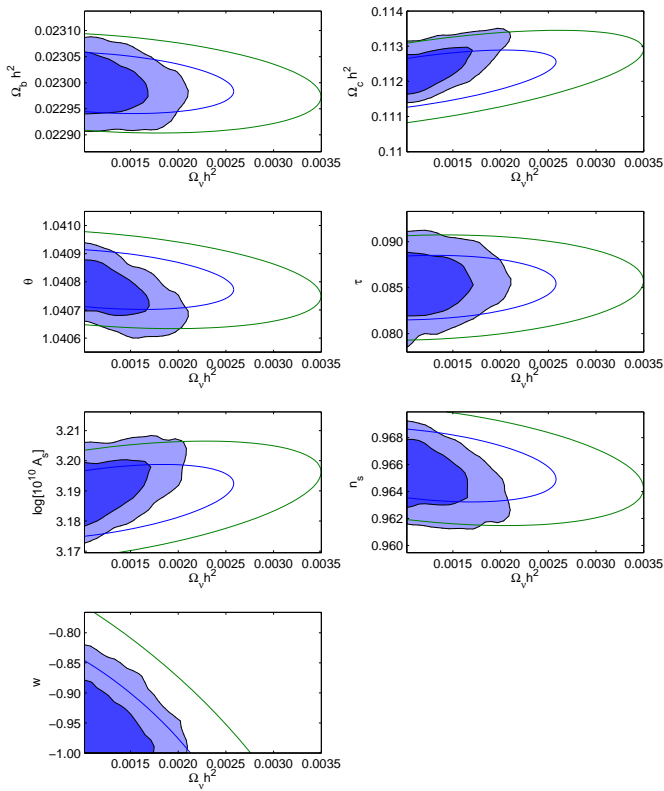}
\caption{Marginalised confidence regions (68 and 95 per cent) between the massive neutrino energy density and the rest of the parameter set for the inverted hierarchy. Shaded regions are MCMC results, and contours are from the Fisher matrix.}
\label{fig:inv_overlay}
\end{figure}

\begin{figure}
\centering
\includegraphics[width=0.5\textwidth]{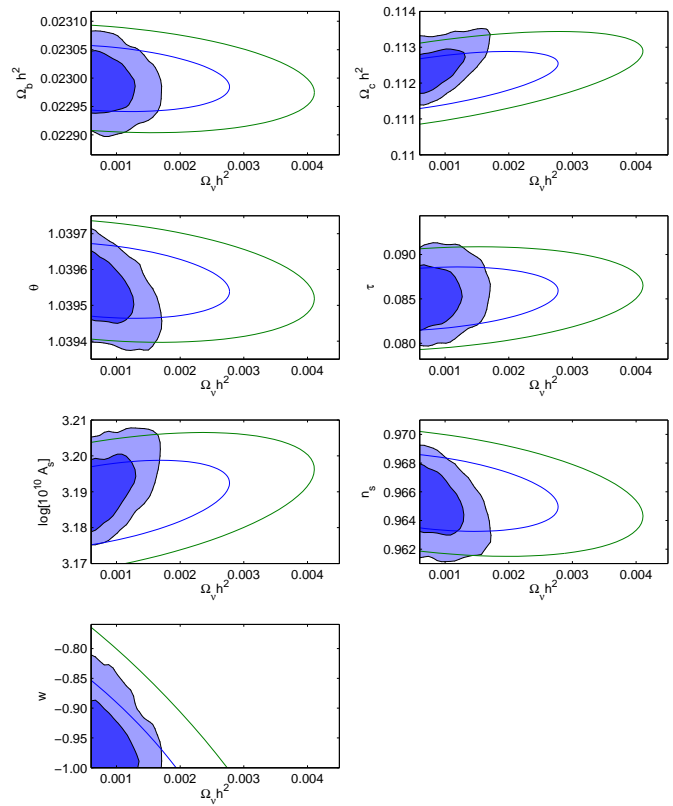}
\caption{As Fig.~\ref{fig:inv_overlay} but for the normal hierarchy.}
\label{fig:norm_overlay}
\end{figure}

We begin by considering CMB data alone. The results of our MCMC runs are displayed in Table~\ref{table:errors}, along with Fisher matrix results for comparison. The forecast 68\% upper limit on the total neutrino mass with CMB T+P+WL with \textit{COrE}-like noise levels is $0.136\,\mathrm{eV}$ for the inverted hierarchy, and $0.098\,\mathrm{eV}$ for the normal hierarchy from the MCMC analysis. Recall that in the fiducial models, $\sum m_\nu = 0.105\,\mathrm{eV}$ in the inverted hierarchy and $\sum m_\nu = 0.062\,\mathrm{eV}$ in the normal hierarchy. For Gaussian marginalised posterior distributions, truncated by the prior on the minimum total mass, these upper limits correspond to $1\sigma$ errors of $0.036$ and $0.039\,\mathrm{eV}$ for the inverted and normal hierarchies, respectively. These results are consistent with the MCMC forecasts in the \textit{COrE} White Paper~\citep{core}. They derived 95\% upper limits on the lightest neutrino mass of the two hierarchies of $0.045\,\mathrm{eV}$ and $0.034\,\mathrm{eV}$ for the inverted and normal hierarchies respectively. Our corresponding values are $0.044\,\mathrm{eV}$ (inverted) and $0.039\,\mathrm{eV}$ (normal), the slight differences probably being due to the slightly different fiducial parameters used. The $1\sigma$ errors on the other parameters are also consistent, with small differences due to the different parameter sets used. The marginalised 68 and 95\% confidence regions for the massive neutrino energy density and the other parameters are plotted in Figs~\ref{fig:inv_overlay} and~\ref{fig:norm_overlay}, along with Fisher matrix results.


Comparing our MCMC results with those from the Fisher matrix, we note significant discrepancies. The Fisher matrix overestimates the forecast $1\sigma$ errors on $\sum m_{\nu}$ and $w$ by a factor of two and on $h$ by a factor of three, but agrees well with the MCMC results for the other parameters. A likely cause of this discrepancy is the strong angular-diameter degeneracy between $h$, $w$, and $\sum m_{\nu}$ in the unlensed CMB power spectra. This degeneracy is not completely broken by the inclusion of the lensing reconstruction and may introduce some non-Gaussianity into the likelihood, thus violating the Fisher approximation.

An additional source of the discrepancy concerns our chosen parameter set. As discussed in Appendix~\ref{app:fisher}, using the ($h$,$\sum m_{\nu}$,$w$) parametrization requires fixing the ratio of certain power spectra derivatives to eliminate numerical noise at intermediate and high multipoles. The results of the Fisher analysis are sensitive to this ratio, which involves $d_A(z_*)$ derivatives. If instead we replace $h$ with $\theta(z_*)$ as a parameter in the Fisher analysis, in which case we enforce $\partial C_l / \partial w = 0$ at intermediate and high multipoles for the unlensed CMB spectra, on transforming back to $h$ we do not recover accurately the original Fisher matrix. Sensitivity to such choices is clearly unsatisfactory, and is the likely explanation for the differences in forecast errors in these parameters. 

A similar discrepancy between MCMC and Fisher forecasts was found in~\citet{perotto_06} in the context of forecasting for \textit{Planck}-like noise, although their disparity is much smaller than ours reported here. In that work, the discrepancies reduced when lensing reconstruction was included. It was argued that, since lensing breaks the main degeneracies in the unlensed CMB, including it brings the posterior closer to a multivariate Gaussian and improves the agreement between Fisher and MCMC analyses. The same is almost certainly true in our forecasts, but is possible that the significantly lower noise levels we have used make the Fisher results more vulnerable to numerical effects even with lens reconstruction included. Note that we checked that running our Fisher analysis with the \textit{Planck}-like noise levels used in~\citet{perotto_06} reproduces their neutrino mass forecasts. It is interesting that the Fisher forecasts on neutrino mass in~\citet{kaplinghat_03} from a rather more sensitive, higher-resolution CMB satellite are comparable to our MCMC results. In particular, they find a $1\sigma$ error of $0.044\,\mathrm{eV}$ assuming two massless and one massive neutrinos.

With CMB T+P+WL, the constraints on neutrino masses are limited by degeneracies with other parameters. To illustrate this, we note that the conditional errors on the total neutrino mass using $\theta(z_*)$ (rather than $h$) in a Fisher analysis are $0.0242$ and $0.0192\,\mathrm{eV}$ for the normal and inverted hierarchies, respectively. In this parameter set, the conditional information on neutrino masses is dominated by $C_l^{dd}$. The most relevant degeneracy for neutrino masses is with the cold dark matter density, as can be seen in Figs~\ref{fig:inv_overlay} and~\ref{fig:norm_overlay}. We may understand this as an effect of the lensing potential, since no such degeneracy is observed with just the unlensed CMB. Inspection of the derivatives of $C_l^{dd}$ with respect to $\sum m_{\nu}$ and $\Omega_ch^2$ reveals broadly similar features. Increasing neutrino mass damps the lensing potential on small scales, with large scales unaffected, as discussed in Section~\ref{sec:background}. Increasing the cold dark matter density boosts the lensing potential on small scales, leaving large scales relatively unaffected. To see this, note that increasing $\Omega_ch^2$ pushes back the epoch of matter--radiation equality to earlier times. For modes that are sub-Hubble during radiation domination, the gravitational potential undergoes oscillations with decaying amplitude until settling down to a constant value again well into matter domination. With matter--radiation equality earlier, the potential decays less during the shorter sub-Hubble radiation-dominated phase and the potential on scales smaller than horizon scale at matter--radiation equality is increased. On large scales there is no such effect. The change in the power spectrum of the gravitational potential appears to dominate other changes, such as the mapping from $l$ to $k$ due to the reduced distance to typical lenses on increasing $\Omega_c h^2$, in determining the effect on $C_l^{dd}$. Since the effect of increases in neutrino mass and cold dark matter density have opposite sign in the lensing power spectrum, the parameters are positively correlated.

Finally we note that there is little difference in our MCMC results between the hierarchies. This is in contrast to the Fisher matrix results, which give a 35\% larger limit on the total mass relative to the fiducial value in the normal hierarchy compared to the inverted. However, since we believe the Fisher results are unreliable, it is clear that this difference is not significant.

\subsection{Inclusion of external datasets}

\begin{table}
  \caption{Upper limits (68\%) on the total neutrino mass $\sum m_\nu$ and dark energy equation-of-state parameter $w$ for inverted (top) and normal (bottom) hierarchies, combining different external datasets (see text for details).}
  \begin{center}
    \begin{tabular}{c c c c c c}
      \hline
      & No priors & $H_0$ & \textit{WFIRST} & BOSS & Combined \\
      \hline
      $\sum m_{\nu} (\mathrm{eV})$ & 0.136 & 0.131 & 0.131 & 0.119 & 0.118 \\
	$w$ & -0.93 & -0.97 & -0.98 & -0.98 & -0.99 \\
	\hline
      $\sum m_{\nu} (\mathrm{eV})$ & 0.098 & 0.095 & 0.095 & 0.082 & 0.080 \\
	$w$ & -0.93 & -0.97 & -0.98 & -0.98 & -0.99 \\	
	  \hline
    \end{tabular}
  \end{center}
  \label{table:priors}
\end{table}

We now consider to what extent external data can improve the constraints on massive neutrinos by breaking degeneracies in the CMB T+P+WL analysis.

\begin{figure}
\centering
\includegraphics[width=0.5\textwidth]{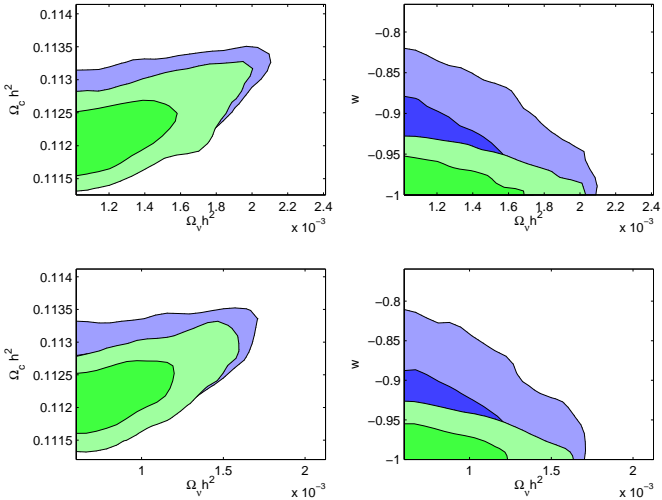}
\caption{Marginalised confidence regions (68 and 95 per cent) between $\Omega_{\nu}h^2$, $\Omega_{c}h^2$, and $w$, with (green) and without (blue) a 2\% prior on $H_0$, for inverted (top) and normal (bottom) hierarchies.}
\label{fig:hubble_prior}
\end{figure}

We start with the effect of including a prior of width 2\% on the Hubble constant. The error on the total neutrino mass (and $w$) is reported in Table~\ref{table:priors}. The Hubble prior changes this little since lensing has broken most of the degeneracy between neutrino mass and $h$ that is present in the unlensed CMB leaving a marginalised error on $h$ from CMP T+P+WL alone comparable to the width of the $H_0$ prior (see Table~\ref{table:errors}). The most important degeneracy for neutrino mass is with the cold dark matter, as seen in Figs~\ref{fig:inv_overlay} and~\ref{fig:norm_overlay}. The inclusion of an $H_0$ prior has little effect on this degeneracy; see Fig.~\ref{fig:hubble_prior}. However, the constraint on $w$ is reduced by over a factor of 2. This is due to the breaking of the degeneracy between $h$ and $w$, a consequence of the geometric degeneracy in the unlensed CMB which is not completely removed by the lensing reconstruction.

\begin{figure}
\centering
\includegraphics[width=0.5\textwidth]{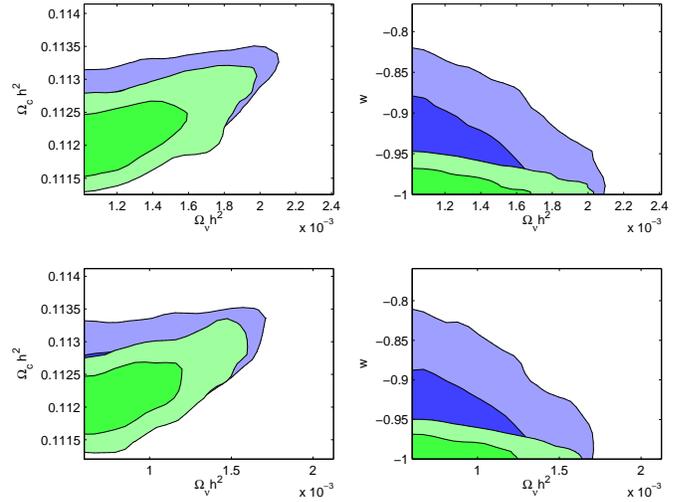}
\caption{As Fig.~\ref{fig:hubble_prior} but with the $H_0$ prior replaced by forecast Type 1A supernovae data from \textit{WFIRST}.}
\label{fig:wfirst_prior}
\end{figure}

When we include forecast observations of Type 1A supernovae from \textit{WFIRST}, we see similar improvements in $\sum m_\nu$ and $w$ as when including the $H_0$ prior; see Table~\ref{table:priors}. After marginalising over absolute magnitude, the supernovae data are essentially distance ratios and so, out to $z=1$, are mostly sensitive in flat models to $\Omega_m$ and $w$. The density parameter is well constrained by supernovae which, when combining with the CMB, sharpens up constraints on $h$ despite the supernovae distance ratios providing no direct measure of this parameter. Moreover, $w$ is well constrained since distances at low redshift are sensitive to the evolution of dark energy. The MCMC results show that the total neutrino mass has little degeneracy with the geometric parameters $h$ and $w$, so the errors on $\sum m_{\nu}$ do not change by much with the inclusion of \textit{WFIRST} data. The degeneracy with $\Omega_{c}h^2$ is again preserved, as shown in Fig.~\ref{fig:wfirst_prior}.


\begin{figure}
\centering
\includegraphics[width=0.5\textwidth]{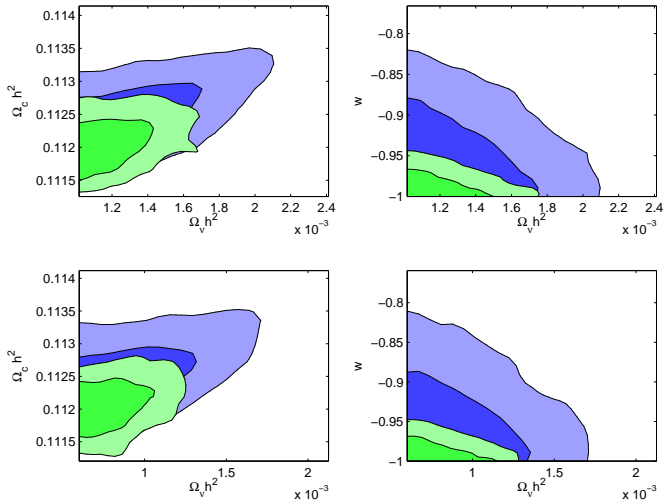}
\caption{As Fig.~\ref{fig:hubble_prior} but with the $H_0$ prior replaced by
forecast BAO data from BOSS.}
\label{fig:boss_prior}
\end{figure}

A greater improvement is seen when forecast BAO data from the complete BOSS is included as a prior. The 68\% upper limits on the summed mass shrink to $0.119\,\mathrm{eV}$ and $0.082\,\mathrm{eV}$ for the inverted and normal hierarchies, respectively, and the upper limit on $w$ becomes $-0.98$ for both hierarchies (Table~\ref{table:priors}). The outperformance of BAO data compared to Type 1a supernovae is probably a combination of effects. Firstly, the supernovae provide only relative distance information (i.e.\ ratios) at low redshifts, which in combination with a well-constrained distance measurement to recombination from the CMB does not provide as strong a constraint on the evolution of the expansion rate as an absolute distance measurement from BAO. This effect is partly compensated by the fact that our BAO measurements are at higher redshifts, closer to the last-scattering surface, and so do not provide as long a lever arm as the supernoave. Secondly, the BAO measurements probe both the angular diameter distance and the Hubble rate, thus offering an internal consistency check that supernoave do not possess. Finally, our forecasted supernovae measurements contain a limiting systematic floor to the error budget which is not shared by the BAO. The effect of BOSS information is shown in Fig.~\ref{fig:boss_prior}. We see from this plot that the degeneracy with $\Omega_{c}h^2$ still remains, although with its strength reduced.

\begin{figure}
\centering
\includegraphics[width=0.5\textwidth]{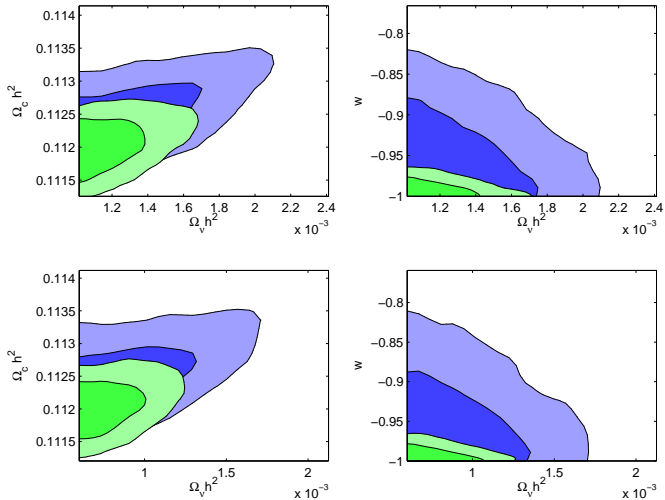}
\caption{As Fig.~\ref{fig:hubble_prior} but with all external data included ($H_0$, Type 1A supernovae, and BAO).}
\label{fig:all_prior}
\end{figure}

Finally, the combination of all priors gives excellent precision on both $\sum m_{\nu}$ and $w$, the 68\% upper limits being $0.118\,\mathrm{eV}$ and $-0.99$ for the inverted hierarchy, and $0.080\,\mathrm{eV}$ and $-0.99$ for the normal hierarchy; see Table~\ref{table:priors} and also Fig.~\ref{fig:all_prior} for confidence regions. The corresponding $1\sigma$ errors assuming truncated Gausssian posterior distributions are $0.018$ and $0.021\,\mathrm{eV}$. The 95\% upper limit in the normal hierarchy is $0.103\,\mathrm{eV}$ (which agrees well with Guassian extrapolation from the 68\% limit). The implication of this is that, even if cosmology had no sensitivity to mass splittings but only to the total mass, the inverted hierarchy would typically be disfavoured at almost the $2\sigma$ level if neutrinos were indeed in the minimal-mass normal hierarchy.


%

%
%

\subsection{Behaviour with fiducial mass}

\begin{table}
  \caption{$1\sigma$ error (see text for details) on the total neutrino mass as a function of fiducial mass, from MCMC runs including only CMB T+P+WL.}
  \begin{center}
    \begin{tabular}{c c c}
      \hline
      Fiducial Mass (eV)& Hierarchy & $\sigma(\sum m_{\nu})$ (eV) \\
      \hline
      0.062 & Normal & 0.039 \\
      0.105 & Inverted & 0.036 \\
      0.140 & Inverted & 0.024 \\
      0.373 & Degenerate & 0.037 \\
      0.559 & Degenerate & 0.037 \\
      \hline
    \end{tabular}
  \end{center}
  \label{table:fid_mass}
\end{table}

We have repeated our MCMC analysis for different fiducial masses. In Table~\ref{table:fid_mass} we report the $1\sigma$ errors on the total neutrino mass for various fiducial values. For the fiducial masses less than $0.373 \,\mathrm{eV}$, the $1\sigma$ error is computed from the 68\% upper limits, assuming the marginalised posterior distribution on $\sum m_\nu$ can be approximated by a Gaussian truncated by the lower limit on the total mass in the respective hierarchy. For total masses greater than $0.3\,\mathrm{eV}$, the presence of mass splitting becomes unimportant, so we forecast assuming three degenerate neutrinos, which speeds up the calculations. In these cases, we simply quote the standard deviation of the samples. The forecasts are for CMB T+P+WL alone.

We observe only a mild variation of the error with fiducial mass, which we interpret as a balance between two competing effects. As the mass increases, the distinctive effect of neutrino free-streaming on the lensing power spectrum is pushed to smaller scales where the reconstruction noise is higher. However, we speculate that the error from the unlensed CMB alone on the total mass should \emph{decrease} until the neutrinos are sufficiently massive to be non-relativistic as recombination, at which point it should flatten out\footnote{For such large masses, the effect on the CMB spectra is roughly linear in the total mass. However, for light masses, their effect on scales $l > 500$ for fixed $\theta(z_*)$ is roughly quadratic in the mass (see Section~\ref{sec:background}) which would give a constant error in the square of the mass.}.

%
\subsection{Hierarchy biasing}

Given our uncertainty in the hierarchy, how should we estimate masses from future data? The Bayesian approach to this problem is that of Bayesian model averaging~\citep{1999STATSCI.14.382H}, in which the posterior probabilities for the parameters in each model are weighted according to the posterior probability of each model. This correctly propagates model uncertainty into parameter errors. In the limit that the evidence for one model is overwhelming, Bayesian model averaging reduces to parameter estimation from that model. In the case that the posterior probabilities for the parameters are similar in each model, averaging is equivalent to using either model alone.

It is therefore interesting to see how the posterior probabilities of the parameters differ between the two hierarchies. To this end, we forecast posteriors adopting the normal hierarchy for fiducial data taken in the inverted hierarchy (with $\sum m_\nu = 0.105\,\mathrm{eV}$. Note that this fiducial total mass can realised in both hierarchies. For CMB T+P+WL, we find no significant bias in any of the cosmological parameters as a result of this procedure. The largest bias is in $\theta(z_*)$, its maximum likelihood value being shifted from its `true' value by $0.76\sigma$. When all priors from external data are included, the greatest bias is in $w$ which is shifted by $0.68\sigma$.

Our findings about bias are in contrast to the results of \citet{debernardis_09}, who found shifts in parameters comparable to their $1\sigma$ errors, with a significant shift in the value of $\sum m_{\nu}$. Direct comparisons are difficult since they use a galaxy weak lensing survey instead of a CMB lensing reconstruction. However, the main source of discrepancy is likely due to the parametrization of the hierarchy in~\citet{debernardis_09}, which was found to be strongly degenerate with $\sum m_{\nu}$.

\subsection{Distinguishing the hierarchies}

Distinguishing the hierarchies is properly a question of model selection. We discuss this for the two cases of fiducial models that are the mimimal-mass normal hierarchy and the minimal-mass inverted hierarchy.

If the masses are in the normal hierarchy, sensitivity to mass splittings is not required to rule out the inverted hierarchy if the true total mass is low enough and the observations have sufficient sensitivity to the total mass. With no sensitivity to mass splittings, the likelihoods for the two hierarchies would necessarily be equal at the same total mass (and all other parameters) but the lower bound on total mass in the inverted hierarchy would exclude the region of high likelihood giving significant odds in favour of the (correct) normal hierarchy. Of course, sensitivity to mass splittings would reduce the likelihood in the inverted hierarchy compared to the normal, further favouring the latter.

We do not consider the minimal-mass normal hierarchy in detail here since our Gaussian-fitting to the likelihood is likely to be very noisy when the prior range excludes the peak of the likelihood. However, we note that lower limits on the Bayes' factor can be inferred from the results of Section~\ref{sec:results}  by assuming no sensitivity to mass splittings. The forecast constraint on total mass using all external data then gives an odds ratio of 15:1 in favour of the normal hierarchy in the most favourable case of the minimum total mass ($0.056\,\mathrm{eV}$). In our forecasting, the peak of the likelihood when analysing with the correct hierarchy is necessarily at the true parameter values. Under one-sigma fluctuations of the peak downwards and upwards, due to cosmic variance and measurement error, the odds ratios vary between 70:1 and 4:1.

\begin{table}
  \caption{Values of $\ln{\langle B \rangle}$ for the inverted to normal hierarchies when the fiducial model is the inverted hierarchy (with $\sum m_\nu = 0.105\,\mathrm{eV}$). The values are for T+P+WL plus the indicated external data.}
  \begin{center}
    \begin{tabular}{c c c c c}
      \hline
      No priors & $H_0$ & \textit{WFIRST} & BOSS & Combined \\
      \hline
      -1.08 & 2.66 & 2.53 & 2.53 & 2.51 \\
      \hline
      \end{tabular}
  \end{center}
  \label{table:bayes}
\end{table}

For the case of the inverted hierarchy, we compute typical Bayes' factors with the approximations discussed in Section~\ref{sec:modelselection}. Our results for $\ln \langle B \rangle$ are shown in Table~\ref{table:bayes}, where the Bayes' factor $B$ is the ratio of evidence for the inverted to normal hierarchies. With no external data, the Bayes factor is negative, indicating that the normal hierarchy would be preferred by the data even though the actual model is inverted. As noted in Section~\ref{sec:modelselection}, this arises because the Occam factor favours the normal hierarchy when the likelihoods have only weak sensitivity to mass splittings and the fiducial total mass can be realised in both hierarchies. We have verified that the other factor in $B$, the ratio of maximum likelihood values, is sub-dominant compared to the Occam factor.


The addition of geometric priors to the CMB data pushes the Bayes' factors into the positive `strong' evidence regime with odds ratios around 12:1 in favour of the (correct) inverted hierarchy. Again, the Occam factor dominates the likelihood ratio (which equals $e^{0.23}$) but now favours the inverted hierarchy, due to small differences in the shapes and peak positions of the likelihoods for the two hierarchies. We note that our results for the Bayes' factors with external geometric data are comparable to those reported by~\citet{jimenez_10} for a full-sky, cosmic-variance-limited galaxy weak lensing measurement with median redshift $3$, although our treatment of the mass hierarchies is rather different to theirs.


Note that to calculate our Bayes' factors, we have to calculate the value of the maximum likelihood when an inverted model is analysed as if it were normal. Since we simply use the maximum-likelihood included in the MCMC chains, there is some error introduced. Since the true value cannot exceed unity, and we typically find minimum $\chi^2$ values around $0.4$, the logarithm of the Bayes' factors could be \emph{lower} by up to $0.2$. One consequence of this is that there is little significance in the anomalously low Bayes' factor found in the `all priors' case.

\section{Conclusions}
\label{sec:conclusions}

We have forecast constraints on cosmological parameters including light neutrino masses with future CMB temperature, polarization and weak-lensing-reconstruction information from a post-\textit{Planck} space-based experiment such as \textit{COrE}. Constraints on mass splittings from oscillation data were used as a prior to delimit two possible mass hierarchies, normal and inverted. We performed MCMC-based forecasts for fiducial models with masses close to the minimal masses in the normal and inverted hierarchies (specifically, we took $\sum m_\nu = 0.062\,\mathrm{eV}$ for normal and $\sum m_\nu = 0.105\,\mathrm{eV}$ for inverted). Our results show that the \textit{COrE} mission alone could place 68\% upper bounds on the total neutrino mass of $0.098$ and $0.136\,\mathrm{eV}$ for the normal and inverted hierarchies. For Gaussian marginalised posterior distributions truncated by the prior on the minimal mass in each hierarchy, these would correspond to $1\sigma$ errors of $0.039$ and $0.036\,\mathrm{eV}$. 

We found good agreement between our MCMC results and a Fisher matrix forecast for all parameters except the neutrino mass, equation-of-state of dark energy $w$ and the Hubble parameter $h$, for which the Fisher matrix overestimates the errors by at least a factor of two. These parameters are highly degenerate in the unlensed CMB spectra for very light masses with only the combination that enters the angular scale of the sound horizon at recombination, $\theta(z_*)$, well constrained. Any independent information in the unlensed CMB spectra on intermediate and small scales constrains combinations of squared-masses as the minimal-masses are approached, these being the leading corrections to the neutrino velocity, energy density and pressure in the ultra-relativistic limit. This dependence on mass further violates the Gaussianity assumption made in Fisher forecasts when the total mass is used as a parameter with ultra-light neutrinos. The $\sum m_\nu$--$w$--$h$ degeneracy is not fully broken by the lensing reconstruction. As a result, our Fisher results proved rather unstable to changes in parametrisation and details of the method of calculation of power spectra derivatives. We therefore advise caution when using Fisher matrices to forecast constraints with ultra-light neutrino masses such as those considered in this work.


We have also considered the addition of forecast geometric information from BAO from the complete SDSS-III BOSS, and Type 1a supernovae distance moduli from \textit{WFIRST}, as well as a future 2\% constraint on the Hubble constant. These are relatively `clean' probes, free from the problems of biasing and non-linearity inherent in using galaxy clustering information directly, and from instrumental effects such as from variations in the point-spread function in cosmic shear. The distance information offered by these probes brings down the 68\% upper limit on the total neutrino mass to $0.118$ and $0.08\,\mathrm{eV}$ for the inverted and normal hierarchies (corresponding to Gaussian $1\sigma$ errors of $0.018$ and $0.021\,\mathrm{eV}$). We also find percent-level precision on the dark energy equation of state parameter $w$. We have found that the BAO data is the most effective geometric probe when combined with CMB T+P+WL. This is likely due to a combination of effects. Compared to supernovae, the BAO measurements provide absolute distance measurements and measures of the Hubble rate. BAO are at higher redshift, which provides less of a lever arm with the angular-diameter distance to last-scattering from the CMB than supernoave, but with the low redshift end not extending so far into dark-energy domination (which confuses neutrino mass determinations from CMB-calibrated relative distances). Moreover, our supernovae forecasts include a limiting systematic floor to the errors which is not the case for BAO.

The precision achieved on the total neutrino mass when all our datasets are combined would be sufficient to disfavour the inverted hierarchy at typically greater than the $2\sigma$ level if neutrinos were in the minimal-mass normal hierarchy. For the alternative case of the inverted hierarchy, we approximated the Bayes' factor with an extension of the `Laplace' method to deal with the prior boundaries. With CMB T+P+WL alone, and a fiducial model with total mass $0.105\,\mathrm{eV}$ close to the minimum value of $0.095\,\mathrm{eV}$, the Occam factor (which gives weight to models whose parameters do not need to be so tightly constrained to fit the data) leads to weak favouring of the (wrong) normal hierarchy over the inverted with 3:1 odds. This situation is overturned by including the geometric datasets, in which case we typically find odds ratios of 12:1 (`strong' evidence) correctly in favour of the inverted hierarchy. The best prospect for distinguishing the hierarchies is for these minimal-mass configurations -- mass splittings and their orderings become irrelevant for cosmological observables at higher masses.


We also ran MCMC analyses to calculate potential biases on parameters by analysing data assuming the wrong hierarchy. We found no biases greater than $1\sigma$ when analysing an inverted model while assuming it to be a normal model. This is in contrast to the results of~\citet{debernardis_09}, although direct comparisons are difficult since they forecast for different datasets and parametrize the hierarchies differently. We note that a straightforward solution to dealing with such potential biases is to perform Bayesian model averaging. This would correctly propagate model uncertainty into parameter errors.

Our results show that CMB lensing, combined with priors on mass splittings from oscillation data and external geometric data, is a promising route to determining whether neutrino masses are hierarchical and, if they are, the ordering of the mass eigenstates. CMB lensing provides a relatively clean measure of the effect of neutrino masses on the clustering of matter below the free-streaming scale. However, even in the most optimistic scenario we consider, the evidence for either hierarchy will never be very strong for these cosmological probes. We would expect our evidence ratios to increase with the addition of other measures of broad-band power in the matter power spectrum, although at the risk of bringing in a host of other systematic effects. One particularly interesting data combination to consider further is combining CMB lensing and cosmic shear tomography. This is free of the issues of galaxy bias, but non-linear effects will be important for cosmic shear and motivates further studies of the effects of individual masses on the non-linear power spectrum~\citep{2012arXiv1203.5342W}.


\section{Acknowledgements}
\label{sec:acknowledgements}

\noindent
AH is supported by an Isaac Newton Studentship from The University of Cambridge, and by the Isle of Man Government. AH wishes to thank Steve Gratton and Lindsay King for helpful advice. We thank Julien Lesgourgues, Laurence Perotto and Martin Bucher for their collaborative work on the \textit{COrE} proposal, on which the current work builds, and Duncan Hanson for computing the noise levels in Fig.~\ref{fig:lensing_noise}

\bibliographystyle{mn2e_fix}
\bibliography{paper_mnras}

\appendix

\section{Protecting degeneracies in the Fisher matrix}
\label{app:fisher}

In constructing the Fisher matrix, we found it necessary to protect several near-exact degeneracies at intermediate and high $l$. We did this for the well-known $\tau$--$ A_{s}$ degeneracy by enforcing a proportionality between the appropriate derivatives. This was also done for $w$--$h$, where the effect on the unlensed $TT$, $EE$, and $TE$ spectra is purely through the angular diameter distance at small scales. The necessity of taking one-sided derivatives for $w$ ($w>-1$) can cause a small phase shift with respect to the $h$-derivative, which can break degeneracies: a simple one-sided derivative at a point actually gives the second-order-accurate derivative about the mid-point. A similar problem was noted in \citet{eht_99} when differentiating with respect to curvature. This makes determining the proportionality constant between the $h$ and $w$ derivatives of the power spectra at high $l$ from the ratio at some intermediate multipole problematic.
Even with a one-sided derivative accurate to second order (using appropriate finite-difference coefficients) this problem did not disappear. Instead, we chose to enforce the degeneracy by computing the derivatives of $d_A(z_*)$ with respect to $w$ and $h$, and using their ratio as the proportionality constant between $\partial C_l/\partial w$ and $\partial C_l/\partial h$ \citep[see Equation B3 of][]{eht_99}. This was done for $l\gtrsim$100, with the precise multipole chosen for a smooth transition. In addition, this enforcement was made across the $TT$, $EE$, and $TE$ spectra by using the same proportionality constant in each case, since the physics governing the degeneracy is in each case the same. Our Fisher results are fairly insensitive to the enforcement of the $h$--$w$ degeneracy, and the choice of multipole ($l\approx$100) above which it is enforced.

No such degeneracy was enforced for $\Sigma m_{\nu}$--$h$ or $\Sigma m_{\nu}$--$w$, since an inspection of the $\Sigma m_{\nu}$ derivative of the CMB spectra at fixed $d_A(z_*)$ revealed a non-zero signal at small scales which did not disappear when more accurate derivatives (4th-order truncation error with large step sizes to beat down numerical noise) were implemented. This signal is the direct contribution of massive neutrinos to the pre-recombination physics discussed in Section~\ref{sec:background}. It is very small, but large enough partially to break the geometric degeneracy with $h$ and $w$.

Double-sided derivatives were used for each parameter except $w$ where the phantom ($w < -1$) regime is not considered and for which we used a second-order-accurate, one-sided derivative. A step size of $5\%$ was used in computing the derivatives, which represents a compromise between numerical noise and Taylor series convergence \citep{methods}. We checked the stability of our results to different step sizes, and found variation of at most 10\% in our marginalised errors with the high accuracy settings used to run \textsc{CAMB}.

\section{Covariance matrix estimation}
\label{app:covmat}

To compare the statistical quantities inferred from MCMC chains to those 
of a Fisher analysis, it is useful to have some idea of the magnitude of 
the statistical fluctuations in the former. For a chain with a large 
number of uncorrelated samples, we would expect these fluctuations to be 
small, but it is important to test whether the scatter is small enough to 
resolve any sharp degeneracies present in the likelihood. In this section, 
we investigate the effect of statistical fluctuations on the 
eigenstructure of the empirical covariance matrix calculated from the chains.


We explore this issue by generating random realisations of a covariance matrix, with the mean matrix given by the output of a Fisher analysis. We use the parameters in Table~\ref{table:fid_model}, but now additionally include the primordial helium abundance $Y_{\rmn{He}}$ as a parameter with fiducial value $0.24$.

\begin{figure}
\centering
\includegraphics[width=0.5\textwidth]{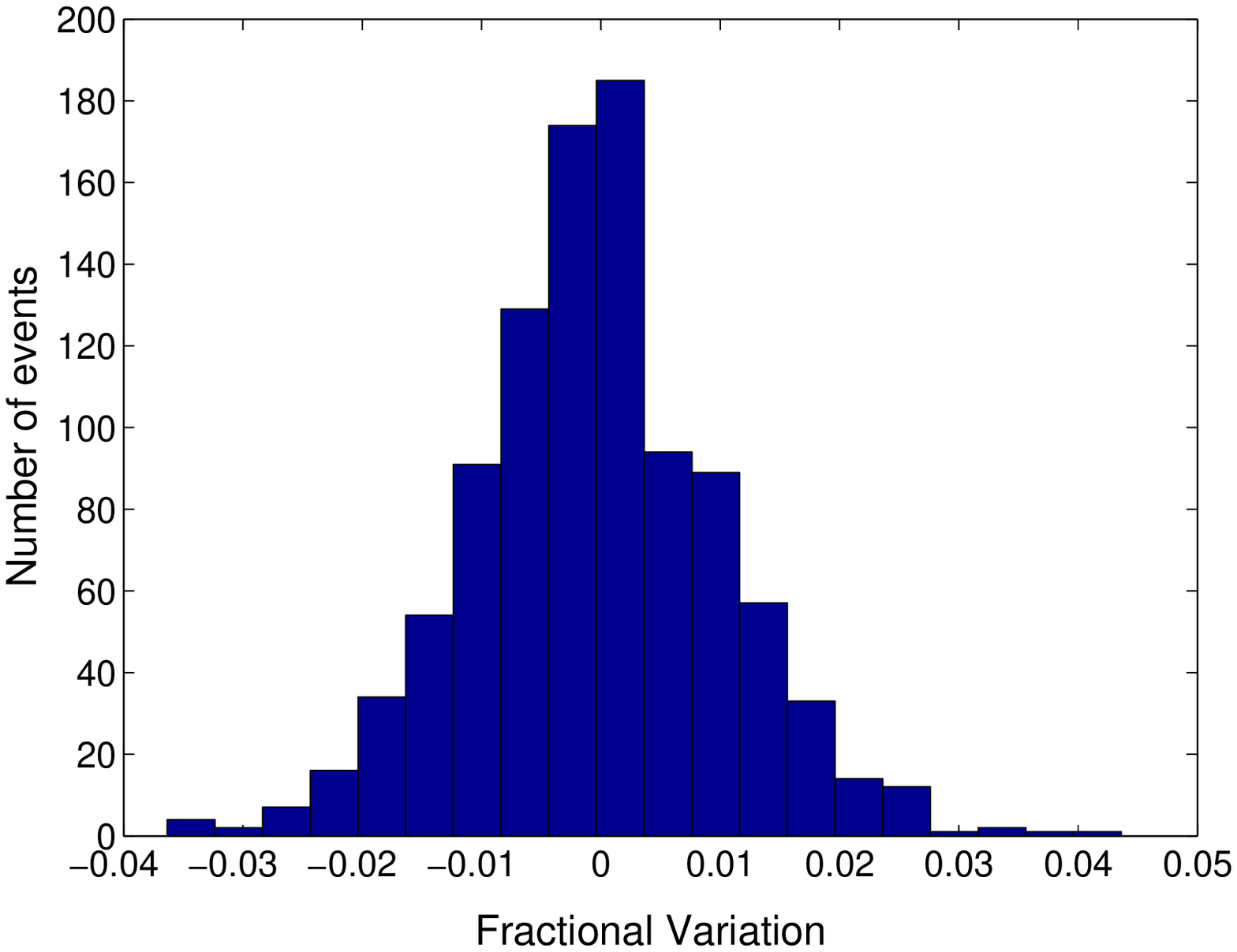}
\caption{Histogram of fractional variations in the lowest eigenvalues of the empirical covariance matrices in an ensemble of 1000 matrices. Each empirical covariance matrix has statistics appropriate to it being estimated from 15000 independent samples from a Gaussian posterior in nine-dimensional parameter space with covariance given from a Fisher analysis.}
\label{fig:rand_eval_lowest}
\end{figure}

\begin{figure}
\centering
\includegraphics[width=0.5\textwidth]{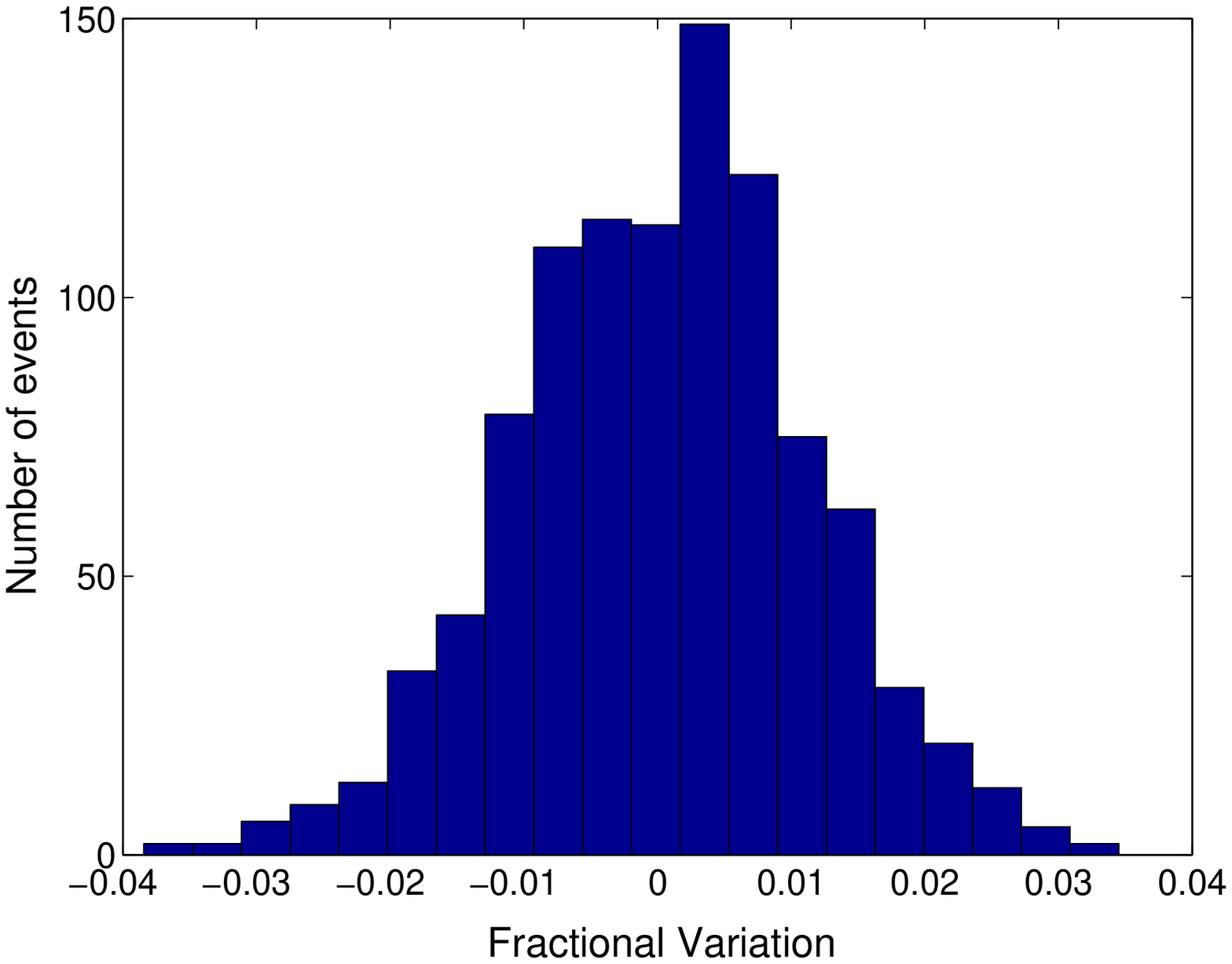}
\caption{As Fig.~\ref{fig:rand_eval_lowest} but for the highest eigenvalue.}
\label{fig:rand_eval_highest}
\end{figure}

\begin{figure}
\centering
\includegraphics[width=0.5\textwidth]{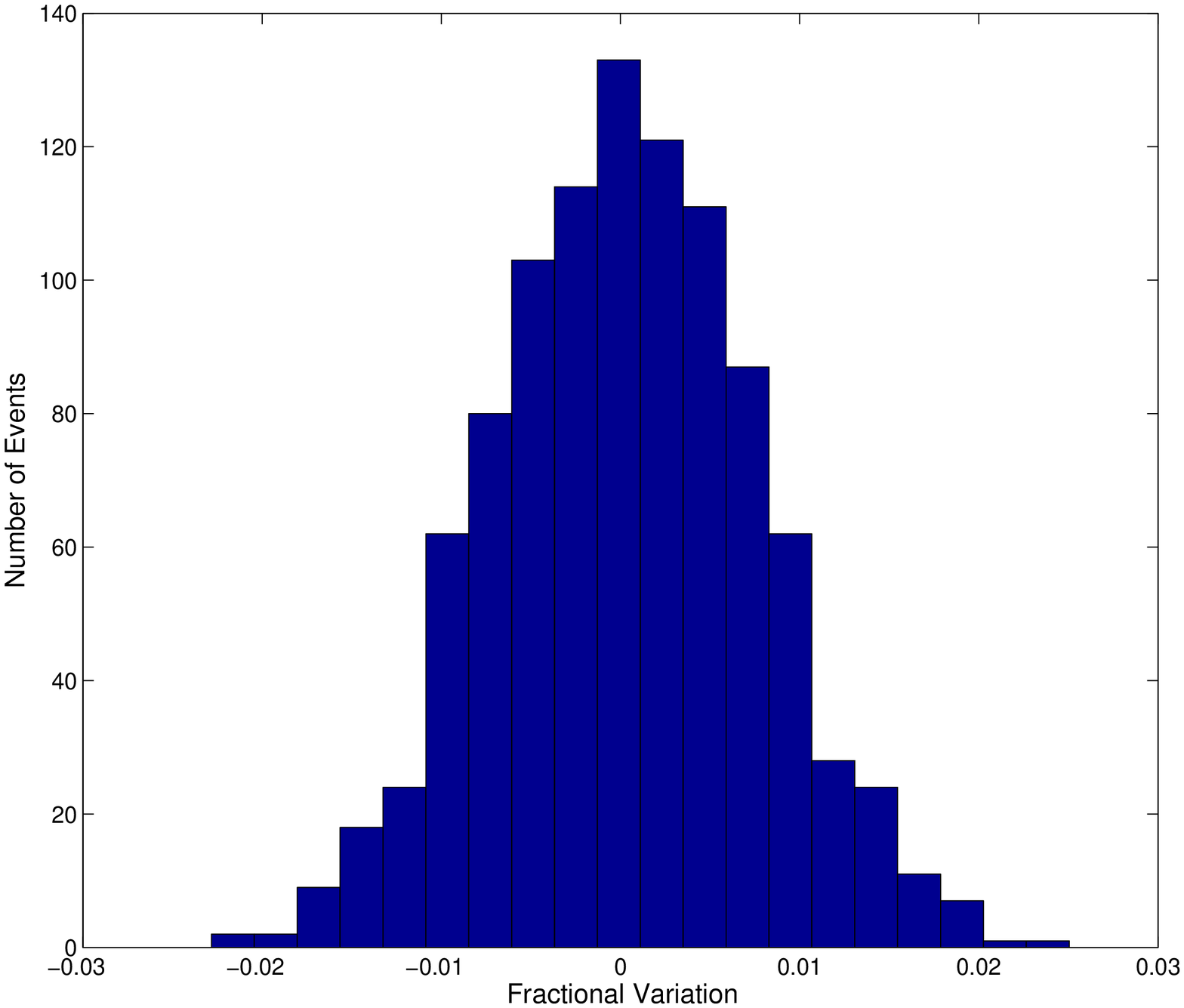}
\caption{As Fig.~\ref{fig:rand_eval_lowest} but for the neutrino mass component of the most poorly constrained eigenvector. This component is the dominant source of the marginalised neutrino mass error in this case.}
\label{fig:rand_nu_angle}
\end{figure}

We assume that each covariance matrix has been estimated from a realisation of $N$ independent parameter samples from a Gaussian posterior, $\{\theta_i^{n}\}_{i=1\ldots 9}^{n=1\ldots N}$. The unbiased estimator for the covariance matrix is then
\begin{equation}
\hat{C}_{ij} = \frac{1}{N-1}\sum^{N}_{n=1}(\theta^n_i - \bar{\theta}_i)(\theta^n_j - \bar{\theta}_j) \, ,
\label{eq:cov_est}
\end{equation}
where $\bar{\theta_i}$ is the empirical mean of the $\theta_i^n$. An application of Wick's Theorem then gives the covariance of these estimators
\begin{equation}
\langle(\hat{C}_{ij}-C_{ij})(\hat{C}_{kl}- C_{kl})\rangle = \frac{1}{N-1}(C_{ik}C_{jl} + C_{il}C_{jk}),
\label{eq:cov_cov}
\end{equation}
where $C_{ij}$ is the `true' covariance matrix of the posterior from which the samples are drawn (we take $C_{ij}$ from the output of our Fisher analysis). We now assume that the estimator $\hat{C}_{ij}$ is Gaussian distributed, which is true for large $N$ by the central-limit theorem. Using our fiducial covariance matrix $C_{ij}$, we draw random realisations of $\hat{C}_{ij}$ from a Gaussian distribution with mean $C_{ij}$ and covariance given by equation~(\ref{eq:cov_cov}), and study the scatter in its eigenstructure about the mean. The results are displayed in Figs~\ref{fig:rand_eval_lowest}, \ref{fig:rand_eval_highest} and~\ref{fig:rand_nu_angle} for 1000 realisations, with $N=15000$. A typical Markov chain might be around 30000 samples long, but since these are correlated, we have implicitly thinned the chain by a factor of two.

Our results show that, in this application, the highest and lowest eigenvalues of the covariance matrix may be determined with percent-level accuracy with $15000$ samples. In Fig.~\ref{fig:rand_nu_angle} we show the distribution of the neutrino mass component of the most poorly constrained eigenvector. In our fiducial model, this eigenvector is responsible for most of the neutrino mass variance. We see that the degeneracy direction most relevant for neutrino mass is constrained at the percent level.

We conclude from this exercise that MCMC covariance matrix estimation is sufficiently robust to the effects of finite sample size for our purposes. It remains to be seen how more degenerate fiducial models may be under-sampled by Monte Carlo techniques. This might be relevant to models with a redshift-dependent dark energy equation-of-state, where we might expect strong degeneracies in the posterior.

\end{document}